\newcommand{\mycomment}[1]{}

\documentclass[%
 aip,
 amsmath,amssymb,
 reprint,%
groupedaddress
]{revtex4-1}
\usepackage{xcolor}
\usepackage{hyperref}
\usepackage{graphicx}
\usepackage{dcolumn}
\usepackage{bm}
\usepackage{natbib}
\usepackage{amsthm} 
\usepackage{subfigure}

\usepackage[utf8]{inputenc}
\usepackage[T1]{fontenc}
\usepackage{mathptmx}
\usepackage{etoolbox}
\makeatletter
\def\@email#1#2{%
 \endgroup
 \patchcmd{\titleblock@produce}
  {\frontmatter@RRAPformat}
  {\frontmatter@RRAPformat{\produce@RRAP{*#1\href{mailto:#2}{#2}}}\frontmatter@RRAPformat}
  {}{}
}%
\makeatother
\begin{document}

\preprint{AIP/123-QED}

\title{Competition for Survival and the Law of Maximum Entropy Production}

\author{Albert Han, Jiri Kataman-Kustwan, Alexey Bezryadin$^*$}
\affiliation{Department of Physics, University of Illinois, Urbana-Champaign}
\email{bezryadi@illinois.edu}

\date{\today}

\begin{abstract}
The law of maximum entropy production (LMEP) is a selection principle stating that a non-equilibrium system will select the path or assembly of paths out of available paths that minimizes the potential or maximizes the entropy at the fastest rate given the constraints. Precise quantitative experimental tests are desirable for LMEP verification. Here we report electrical-circuit experiments on suspensions of silver particles which self-organize under electric field, form conducting paths and thus generate entropy by Joule heating. Our experiments on parallel-connected pairs of samples demonstrate a competitive relationship, thereby supporting the selection of the most efficient path. We find: (1) With sufficient voltage, a conducting bundle of silver particles forms, which acts as a self-organized dissipative structure (SODS). (2) With two samples in parallel, usually only one can develop SODS, due to the competition effect. As one sample approaches the maximum EPR, the others EPR is near zero. This confirms the path selection principle: only the highest EPR path survives. (3) At the end of the evolution, the sum of the sample EPR and the power supply circuit EPR approaches the calculated maximum. The deviation from the maximum is slightly lower when there are two competing samples. We identify the selection mechanism enabling natural selection and LMEP: A SODS gains a form of physical "fitness" when it establishes a pathway that efficiently suppresses the driving thermodynamic potential. As it develops and approaches the maximum EPR, the SODS modifies its nonequilibrium environment by consuming available free energy and eliminating the gradients that could support alternative pathways. We also examine the global implications of LMEP and propose that it serves as a driving mechanism propelling the hypothetical ascent of civilizations along the Kardashev scale.

\end{abstract}

\maketitle 

\section{Introduction}

Thermodynamics of equilibrium systems is well understood and has a governing extremum principle, namely, the second law of thermodynamics. However, the behavior of systems far from equilibrium remains mysterious and a subject to active debates\cite{MartinezKahn2010}. One thing is clear--complex systems far from equilibrium frequently show self-organization phenomena which accelerate entropy generation. The most famous example is the Benard convection cell\cite{Jia2014} which is a self-organized dissipative structure (SODS), the function of which is to increase the entropy production rate (EPR)\cite{Martyushev-2006,Martyushev-2025}, which is realized  by convection. One can argue that the life itself is a non-equilibrium phenomenon or SODS\cite{Schrodinger1944}.

A general concept that has attracted considerable attention in recent decades is that nonequilibrium systems may self-organize in ways that enhance and even maximizes the rate of heat dissipation and EPR, subject to the constraints imposed on the system \cite{Lotka-1945,Lotka1922a,Lotka1922b,Swenson1989,Bezryadin-1999,kleidon_2010, Swenson2025,bezryadin-2015}. Such entropy production maximization ideas have been explored across a wide range of fields, including climate science \cite{Paltridge-1975,Paltridge-1978}, biology \cite{EnzymeCascadeDobovisek}, and physics \cite{Martyushev2021-Review}. Applications and proposed examples include the Earth's climate, biological systems, self-organized electrorheological fluids \cite{Bezryadin-1999}, and many other nonequilibrium systems \cite{Kleidon2009-fg}. However, the precise formulation, theoretical status, and general applicability of entropy production maximization ideas as potential fundamental principles remain subjects of ongoing debate.

Understanding physical mechanisms leading to biological organisms and their competitive and/or collaborative behavior remains a fundamental problem of modern physics\cite{schrodinger1974}. To explain and predict selection processes from a physics prospective, Alfred J. Lotka proposed\cite{Lotka1922a,Lotka1922b} a maximum power principle, according to which natural selection tends to favor organisms and systems that maximize useful energy flux (power) through themselves, insofar as competing alternatives are available. Lotka wrote that natural selection would favor those organisms that could capture and utilize more available energy. Later, Odum and Pinkerton propose the following formulation\cite{Odum1955}: Under the
 appropriate conditions, maximum power output is the criterion for the
 survival of many kinds of systems, both living and non-living. Since the entropy production is related to the dissipated power, Lotka's maximum power principle is related to maximum entropy production because increasing the throughput of available energy in irreversible systems generally increases the rate of entropy production\cite{Odum1955,Hill1990,Odum1988}. Sciubba as well as Vallino and Algar analyzed the relationship between Lotka's principle and the entropy production maximization. \cite{Sciubba2011,Vallino2016}

A general formulation of the Law of Maximum Entropy Production (LMEP) was proposed by Swenson \cite{Swenson1989,Swenson2010}. The LMEP is advanced as a general evolutionary principle governing the organization and development of physical, biological, and ecological systems \cite{Swenson1989,Swenson2025,Wang2026}. According to LMEP, nonequilibrium systems select, from the available possibilities, the pathway or combination of pathways that dissipates the available potential and produces entropy at the maximum possible rate, subject to the prevailing constraints \cite{Swenson2025}. In this perspective, the emergence of organized structures is not opposed to the second law of thermodynamics; rather, the formation of self-organized dissipative structures (SODS) facilitates more effective energy and matter flows, thereby increasing the overall rate of entropy production. For an interesting concrete example, see Ref.\cite{Gimsa2020} where LMEP successfully describes the field-induced orientation of ellipsoids of rotation, which is a system similar to the scope of our experiments. 

It is important to distinguish LMEP from the recent suggestions such as the Maximum Entropy Production Principle (MEP)\cite{kleidon_2010,Kleidon2009-fg,Dyke-2010-Theory} introduced to describe self-organization phenomena in non-equilibrium thermodynamics. The MEP principle posits that open, non-equilibrium thermodynamic systems with sufficient degrees of freedom will naturally evolve toward a steady state that maximizes the production of entropy. The MEP can be understood as a selection criterion for determining the steady state or the distribution of fluxes among alternative non-equilibrium states, given specified constraints. On the other hand, LMEP places greater emphasis on the path selection, i.e., on the evolutionary dynamics through which systems select and develop pathways that enhance their capacity for entropy production at all times. Thus, whereas steady-state MEP applications often focus on the selection of the final steady state, the LMEP emphasizes the selection of the pathways through which a system evolves. LMEP is the local path selection by steepest rate law, that applies universally across all ranges, near and far from equilibrium.\cite{Swenson1989, Gimsa2020}
On the other hand, the MEP describes stationary states in non-equilibrium processes.\cite{Bradford-2013,Martyushev2021-Review, MARTYUSHEV201417, Hall-2023, Ozawa2003-kg, Bradford-2013, Dyke-2010-Theory}

A mathematical evolution model, which is simple but general, has been proposed in Ref.\cite{Bezryadin-1999} In this model, a self-organized dissipative structure (SODS) is presented as an ensemble of parallel self-organized dissipative elements (SODE). The key assumption of the model is that each SODE (which can be exemplified by a conducting chain composed of metallic particles) is destroyed (due to e.g., Joule heating) if the power dissipated in it increases above a fixed threshold. In such formulation, each new SODE, appearing within SODS, increases the entropy production rate. The main prediction of the model is as follows: The behavior of a SODS (which is an ensemble of SODE) changes qualitatively once the power dissipated within the entire SODS reaches its theoretical maximum, set by the constraints (available potentials and limiting factors). This condition defines a certain critical point for the system behavior. At this critical point, a transition from an avalanche regime to an avalanche-free regime takes place. This critical point is such that the dissipated power inside SODS is at its maximum while the total dissipated power, in the SODS and in the (electric) potential sources taken together, is at 50\% of the maximum possible value. The model predicts that the evolution does not stop at this critical point. The second stage of the evolution (avalanche-free regime) is characterized by a reduction of the power dissipated inside SODS, while the total dissipated power (SODS+free energy source) continues to approach its theoretical global maximum. It is understood that the entropy production rate equals the dissipated power divided by the temperature of the environment.\cite{Bezryadin-2016} According to the model, the evolution does not stop as the entropy generated within self-organized dissipative structure equals its theoretical maximum. The end point of the evolution is the maximum entropy production rate (or maximum dissipated power) in the entire circuit including SODS as well as the energy source. This work suggests that any physical principle of maximum entropy production must specify if the entropy under question is the one generated within self-organized system itself or within the energy source producing non-equilibrium conditions or the total entropy of all components involved, including the energy sources. The behavior predicted by this model is in agreement with high precision measurements performed on carbon nanotube suspensions\cite{Bezryadin-1999}.  

The ideas of the maximization of the entropy production rate have been applied for problems in atmospheric science\cite{Paltridge-1975,Paltridge-1978,Lorenz2001-ae}, mantle convection\cite{Lorenz_2002}, and fluid flow \cite{Martyushev2007-ab}. 
One notable application of the maximum entropy production ideas includes the calculations of Paltridge, who showed that maximizing the entropy production in Earth’s atmosphere produces temperature distributions that is in an approximate agreement with the measured values.\cite{Paltridge-1975,Paltridge-1978} Similar calculations have been done for the atmospheres of other planets like Mars and Saturn’s moon Titan \cite{Lorenz2001-ae}. Maximization of the entropy generation has also been used to address other problems in Earth science, such as mantle convection \cite{Lorenz_2002} and thermal circulation in the ocean \cite{Shimokawa2002-ue}. In physics, the maximization of the entropy production has been applied to problems such as fluid flow\cite{Martyushev2007-ab}, and Rayleigh-Benard convection\cite{Bradford-2013} and self-organized motile ("wiggling") clouds of carbon nanotubes\cite{bezryadin-2015}. In biology, these ideas have been used to model photosynthesis\cite{JURETIC-2003} and biological evolution \cite{Lotka-1945}. 

However, despite these interesting examples, it is still a heavily debated field of research, and not generally accepted due to a lack of a high-precision, reproducible experimental confirmations. There are publications to disprove the MEP \cite{Andresen1984-fm, Ross-2012, Polettini-2013}, while proponents argue that the counterexamples provided are outside of the MEP’s applicability\cite{MARTYUSHEV201417}. These issues are complicated by conflicting formulations of the MEP. Some formulations maximize the global dynamic steady state entropy production rate, while others require local maximization of the EPR.\cite{Bradford-2013,Martyushev2021-Review, MARTYUSHEV201417, Bezryadin-2016, Hall-2023, Ozawa2003-kg, Bradford-2013, Dyke-2010-Theory} Generally speaking, despite multiple applications, outlined above, the maximization of the entropy production is still not well understood or widely accepted. This is due to the fragmented formulations, not specifying which entropy is considered (within the system or within the energy source or the total), and a lack of high-precision and reproduced laboratory experiments testing this principle.

An important question is whether nature performs a
``selection'' solely during biological evolution or whether it
universally operates such a ``selection'' mechanism across all non-equilibrium physical
processes, namely physical selection \cite{SwensonTurvey1991, Swenson2010, Lotka-1945}. If one assumes
that nature universally selects all physical processes, then according
to Occam's Razor, these processes should adhere to a unified principle. Otherwise, we would be obliged to hypothesize the existence of two distinct yet non-contradictory physical laws. Given the usefulness of the second law of thermodynamics and other extremum principles in physics, it is natural to look for an analogous principle that governs non-equilibrium systems, as many phenomena of great interest take place far from equilibrium\cite{,  Saglam2022, Morowitz2007, Attal_2021,https://doi.org/10.1111/ecin.13159,Swenson1998, DEMETRIUS20001,
Dewar_2005, Lotka-1945, Swenson2020, 10.1098/rsta.2022.0277,Lotka1922a,Lotka1922b,Onsager-1931,Onsager-1931-2,Ziegler-1962, Bezryadin-1999, Martyushev-2006, Martyushev2007-ab, kleidon_2010}.

The hypothetical maximum entropy production approach, like any fundamental concept in physics, must be validated through a series of precise, reproducible experimental measurements—a critical area that remains insufficiently developed. The lack of high-resolution quantitative measurements of the entropy production in non-equilibrium systems capable of self-assembly has hindered empirical support for the MEP principle, leaving it somewhat philosophical. In many natural processes, the system is too complex to calculate entropy production analytically or to measure EPR with high accuracy. Also, it is usually difficult to calculate the maximum possible entropy production rate, which is needed to be known to confirm that idea.  

In an effort to overcome these difficulties, electrical circuit experiments with suspensions of electrically conducting carbon nanotubes have been performed.\cite{Bezryadin-1999} In such experiments, the Joule heating and thus the EPR can be measured using high-precision equipment. The experiments revealed that under non-equilibrium conditions created by an applied voltage the system repeatedly generates conducting chains or paths and eliminates overheated pathways through avalanche dynamics. Each conducting chain is described as a self-organized dissipative element (SODE). Ultimate stabilization occurs when the conducting network develops to the point where energy throughput becomes sufficiently high that power dissipation within the energy-supply circuit exceeds that within the SODS (the entire ensemble of SODEs), per se. Also, previous investigation on artificial non-biological systems have demonstrated that self-organized dissipative structures can spontaneously move in a coordinated, exhibiting animal-like motility.\cite{bezryadin-2015} Also, a "training effect" has been previously observed in such SODS\cite{Bezryadin-2016}.

Here, our goal is to investigate evolution stages and selection mechanisms using a model system based on a suspension of silver particles. The key advantages of this model system is that (a) the dissipated power and thus the entropy production rate can be measured with high precision and (b) the theoretical maximum of the dissipated power and the entropy production rate can be easily calculated from the electrical circuit diagram. Note that factors (a) and (b) are needed for a convincing testing of any maximum entropy production model. Our system exhibits a compelling self-assembly behavior and enables precise quantification of the heat generated and the EPR (entropy production rate). This work explores the possibility that selection is not an exclusive attribute of life but a universal principle governing the evolution of non-equilibrium systems. 

Our general conclusion is as follows: If driven non-equilibrium systems continually generate competing dynamical configurations (SODS) then only those which are able to effectively \textit{reduce} their thermodynamic free-energy-providing gradients can persist and survive, provided that they do not destroy itself by the amount of heat produced. Then, biological natural selection may represent a particular realization of a more general process of physical selection. In this context, physical selection provides a dynamical link between microscopic fluctuations and macroscopic thermodynamic organization: it is through the continual generation, competition, and stabilization of alternative states that systems discover pathways of increasingly effective energy dissipation. Our experiments show that the necessary feedback mechanism arises from the fact that more efficient dissipative structures reduce the thermodynamic (e.g., electric) potentials available to competing dissipative structures. Consequently, dissipative structures with the highest rates of entropy production become stabilized and dominant, in accordance with LMEP.

Although entirely non-biological, the reported behavior exhibits the essential ingredients of an evolutionary process—variation, competition, persistence, and selection—providing evidence that selection may be a fundamental organizing principle of non-equilibrium physical systems.

The silver particles used in our experiments, either Ag nanowires or silver flakes, are suspended in isopropanol and subjected to a strong electric field by applying a high voltage across a pair of metallic electrodes immersed in the suspension. The non-equilibrium condition is due to this voltage applied, which represents a gradient of the electric potential. The voltage and the current are precisely measured during the experiments. The main effect is that, once a conducting chain of Ag particles forms, it reduces the electric potential gradient (i.e., the voltage drop across the suspension), thereby depriving competing structures of access to the source of free energy. This appears to be the mechanism of the observed competition effect. We also analyze the effect of competition on the global entropy production rate of the entire system. 

To isolate the competition effect from other collective self-assembly phenomena, our setup involves two self-assembly subsystems (two samples) arranged in a parallel electrical connection within the circuit. By measuring the current $I$ and the voltage $V$ within each sample, the Joule heating is calculated as $P=IV$ and the entropy production rate (EPR) as $dS/dt=IV/T$, where $t$ is time and $T$ is temperature.

The competition effect is the main result reported here. Systematic experiments show that only one of the two parallel samples achieves the maximum contribution to the global entropy production rate. The other sample drops to the minimum EPR. Our results demonstrate that if resources are limited the competition makes it improbable for all subsystems to succeed and contribute their potential maximum EPR to the global EPR. Thus we confirm that the most efficient entropy-producing path is selected\cite{Swenson1989}. 

We propose the underlying physical mechanism which explains the selection originally postulated by LMEP\cite{Swenson1989}: A self-organized dissipative structure gains a form of physical "fitness" when it establishes a pathway that efficiently channels available energy flows. As it develops, it modifies the non-equilibrium environment by consuming gradients and reducing the available free energy that could support alternative pathways. This creates a selection mechanism: configurations capable of maximizing the entropy production persist, while competing configurations are suppressed because the thermodynamic resources necessary for their emergence have been diminished.

Also, we report two other observation which can potentially be generalized to other types of SODS: (1) It is found that, if competition is present by design, the global EPR is slightly higher in comparison to analogous experiments but without multiple competing samples. (2) Our experiments demonstrate that initially, most of the heat (and entropy) is generated inside the SODS because its electrical resistance, $R_{sample}$, is initially much higher than the resistance of the voltage/current supply circuitry, defined by $R_{series}$. Yet, at the second (final) evolution stage, at which SODS consumes almost all available power, most of the heat (and entropy) production shifts into the energy supply circuitry (modeled by a fixed resistor $R_{series}$ connected in series with the voltage source). If generalized, this observation parallels the Kardashev scale, which is a framework in astrophysics and that ranks civilizations by how much energy they can harness and use — because, within this frame, the energy consumption outside of biological organisms themselves is the key parameter characterizing technological power of a civilization. (3) We conclude that the competition effect is the cause of the natural selection. The competition and selection can be understood as a shift of the driving thermodynamic gradients or potentials away from the self-organized dissipative structure (SODS). Such shift suppresses the development of competitor SODS. Thus the SODS with the highest EPR is naturally selected, in accordance with LMEP.\cite{explanation}

\section{Experimental Setup}

The setup is similar to previous experiments\cite{Bezryadin-1999,bezryadin-2015}. In this section, first we describe the sample preparation procedures and then discuss the circuit used to perform high-precision electrical measurements. Isopropanol was used as the solvent in all reported experiments. Novarials A100UL silver (Ag) nanowires (100nm diameter, about 100$\mu$m long) and Alpha Aesar 80$\%$ < 20 micron Ag flakes were used to form electro-rheological suspensions. Suspensions of Ag nanowires between $0.025$ and $1.5 \frac{mg}{ml}$ were tested, and suspensions of Ag flakes between $0.3$ and $3\frac{mg}{ml}$ were tested.

To prepare each sample, approximately 10 ml of solution was placed into a glass vial equipped with two stainless steel electrodes, $0.5$mm diameter, $20$mm length. The electrodes, in the shape of needles, were inserted through a rubber or plastic cap into the solution. The distance between the electrodes was $6$ mm. The electrodes allowed the samples to be connected to a voltage source. The voltage source was connected in series with the resistor to model the source's internal dissipation and limit the power that can be drawn from it. Such limitations mimic real-life situations in which resources are limited and the EPR cannot be set to infinity. 

Before and between experiments, Ag nanowires were suspended in isopropanol by shaking the container until the solution was visibly homogeneous. The Ag flake samples were sonicated until visibly homogeneous. We have noticed that ultrasound sonication can break nanowires and change the samples qualitatively. Thus ultra-sonication was not used to re-agitate samples with Ag nanowires. The concentrations tested were chosen to ensure that significant self-assembly was possible, i.e, a self organized dissipative structure (SODS) could form between the electrodes. Yet, the concentration was low enough so that the current at zero time ($t$=0) is much lower than the maximum possilbe current $I_{max}=V_{source}/R_{series}$. 

First, measurements were performed on single samples to verify that these Ag particles (either Ag nanowires or Ag flakes) were able to self-assemble under the present conditions, similar to previously observed self-assembly in single-wall carbon nanotube suspensions \cite{Bezryadin-1999,bezryadin-2015,Bezryadin-2016}.  

\begin{figure}
    \centering
    \includegraphics[width=1\linewidth]{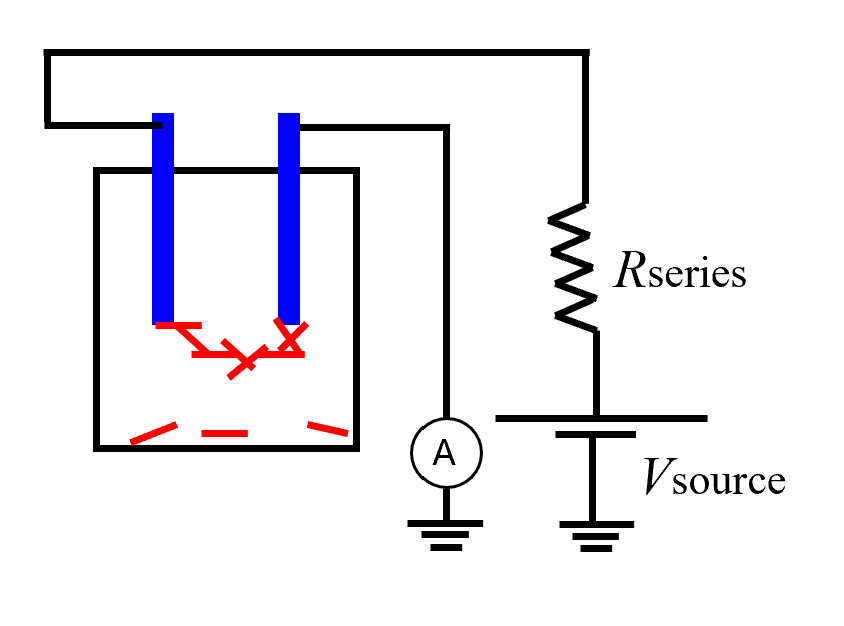}
    \caption{Single-sample measurement setup. The container with Ag nanowire suspension in isopropanol is shown on the left. It is equipped with two electrodes (blue), submerged into the fluid, with Ag nanoparticles (either nanowires or micron-scale flakes). The fluid is placed in a glass container (black rectangle). The nanowires (red) are shown schematically to self-assemble on the electrodes, by forming an electrically conducting chain, which is referred to as a self organized dissipative structure (SODS), since it dissipates electrical energy of the voltage source into heat. The resistor $R_{series}$ limits the current in the circuit. It sets an upper limit for the dissipated power. The current in the circuit is measured with a high precision  ammeter "A".}
    \label{fig:Onechannel}
\end{figure}

\begin{figure}
    \centering
    \includegraphics[width=1\linewidth]{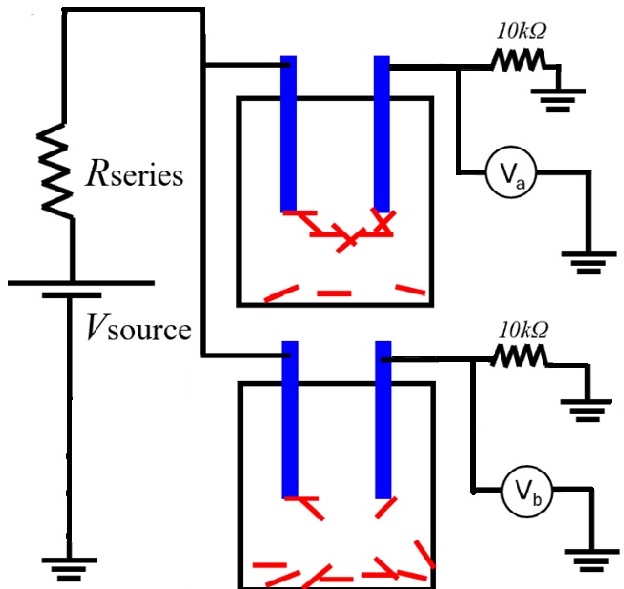}
    \caption{
    Two-sample measurement setup. The two containers, containing Ag micro-flakes suspended in isopropanol, are connected in parallel. The setup is designed to model the natural competition for resources. In this case, if a SODS forms in one sample (the top one in this example), the voltage across both samples drops, and the second sample cannot form a dissipative structure because the electric field becomes too weak. The voltages $V_a$ and $V_b$ are measured with voltmeters (PicoScope) and then used to calculate the currents through the samples. In this case the current in sample "A" is $V_a/10k\Omega$ and current is sample "B" is $V_b/10k\Omega$. Note that $V_a$ and $V_b$ are not voltages on the samples but they are voltages on small series resistors used to measure the currents.  We always ensure that $R_{series}\gg 10k\Omega$.}
    \label{fig:twochannels}
\end{figure}

A schematic of our circuit is shown in Fig.\ref{fig:Onechannel}. A bias voltage ($V_{source}$) is provided by a Keithley 6517B. Voltages between $5-30V$ and $40-300V$ were tested for Ag nanowires and Ag flakes respectively. A series resistor ($R_{series}$) was placed in series with the sample to limit the current. Various series resistors were tested, in the range between $5k\Omega$ and $500k\Omega$.

\begin{figure}[t]
    \centering
    \includegraphics[width=1\linewidth]{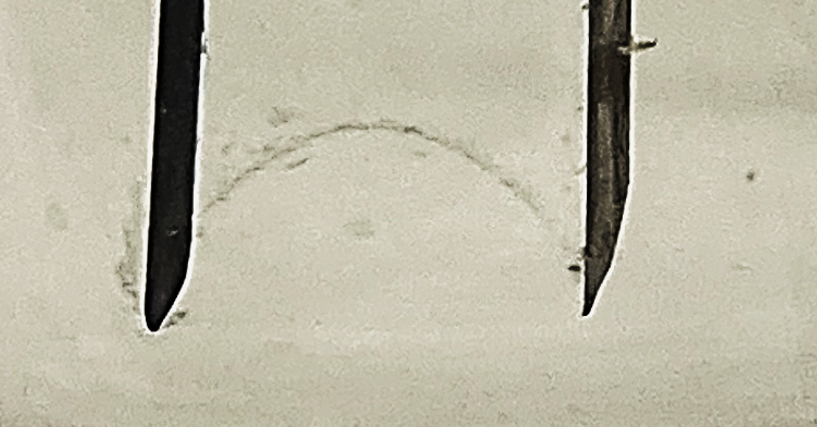}
    \caption{Formation of a self organized dissipative structure (SODS), namely a bundle of Ag nanowires in isopropanol (sample S1). The suspension concentration was $0.025 \frac{mg}{ml}$. The Ag nanowires were 100 nm in diameter and up to 100 microns long. The sample, namely the two stainless steel needles immersed in the suspension, was connected in series with a 15.9k$\Omega$ resistor and biased with $V_{source}=$18V. The image shows the Ag nanowire bundle formed at $t\simeq$4000s. A small, convex, arched Ag chain (the SODS), curved upwards, has formed, which connects the two electrodes (needles).}
    \label{fig:Agnanowirechain18V}
\end{figure}

In Fig. \ref{fig:Onechannel}, "A" represents the ammeter (Keithley 6517B), which can measure currents with high precision, down to the level of a few fA ($10^{-15}$A).   In most experiments, self-assembly was observed and the entropy production increased drastically. The SODS is an electrically conducting chain of silver particles (nanowires or flakes), schematically shown in Fig.\ref{fig:Onechannel} as a chain-like structure (see the red lines). This  chain structure forms between the electrodes (blue) due to the effect of the applied voltage and the corresponding electric field, which polarizes the nanowires (red). An example of such Ag nanowire chain (DODS) is shown in Fig.\ref{fig:Agnanowirechain18V}.

The SODS (the Ag particles chain) is electrically conducting, so that its formation drastically decreases the resistance between the electrodes (blue). Thus, it increases the Joule heating power $P_{global}=V_{source}I$ in the circuit, and, correspondingly increases the total (global) entropy production rate (EPR), $dS_{global}/dt=P_{global}/T$. Here $I$ is the current in the circuit, $V_{source}$ is the voltage provided by the voltage source, $T$ is the temperature of the environment, and $t$ is the time. 

The global entropy production rate is the sum of the entropy produced due to the heat released within the sample and the heat leased in the energy supply circuit, i.e., the series resistor $R_{series}$. The experiments are conducted at room temperature, so $T=295$K. Therefore the total entropy production is 

\begin{equation}
    dS_{global}/dt=IV_{source}/T 
    \label{eq:GlobalEPR}
\end{equation}
The total Joule heating and the total entropy production rate of the circuit is directly proportional to the current, since the source voltage is fixed. Therefore, the maximum entropy production state occurs when $I$ is largest, or the total resistance of the circuit is smallest. The calculated absolute maximum (CAM) of the total dissipated power, $P_{max,global}= V^2_{source}/R_{series}$, and the corresponding calculated maximum global entropy production rate is
\begin{equation}
 \max(\frac{dS_{global}}{dt}) = \frac{1}{T}\frac{V^2_{source}}{R_{series}}
    \label{eq:GlobalEPRMax}
\end{equation}
Note that this EPR cannot be achieved since it would require the resistance of the SODS (Ag chain) to become zero, which is not realistic. However, the sample can always evolve closer and closer to this maximum since more Ag particles can join the chain and make it more and more conducting over time. Such evolution process is frequently observed in the experiments.

The power dissipated within the sample only is $P_{sample}=IV_{sample}$. In our experiments the voltage of the source, $V_{source}$, and the current, $I$, is measured as a function of time.  The voltage drop on the resistor, $IR_{series}$, is calculated by Ohm's law, since the series resistor value is know. Thus the voltage on the sample is $V_{sample}=V_{source}-IR_{series}$.

The entropy production rate by the sample itself (sample EPR) is $dS_{sample}/dt=P_{sample}/T$. Interestingly, this sample EPR can (and frequently does) achieve its theoretical maximum. This happens under the resistance matching condition, namely when the resistance of the dissipative structure ($R_{sample}$) equals the resistance of the series resistor, i.e., when $R_{sample}=R_{series}$, in which case $V_{sample}=V_{source}/2$. At this point the power released in the energy supply circuit (the resistor) equals the Joule power released in the SODS. Note that $V_{sample}+V_{resistor}=V_{source}$, where $V_{resistor}=IR_{series}$ is the voltage drop on the resistor. The maximum power which can be dissipated within the sample is $P_{max,sample}=(V_{source}/2R_{series})(V_{source}/2)$. The corresponding maximum value of the EPR within the sample is

\begin{equation}
 \max(\frac{dS_{sample}}{dt}) = \frac{1}{T}\frac{V^2_{source}}{4R_{series}}=\frac{1}{4}\max(\frac{dS_{global}}{dt})
    \label{eq:SampleEPRMax}
\end{equation}

In the the general case, the formula for the entropy production by the sample equals the Joule power dissipated by the sample divided by the temperature of the environment 
\begin{equation}
    dS_{sample}/dt=R_{sample}[V_{source}/(R_{series}+R_{sample})]^2/T
    \label{eq:SampleEPR}
\end{equation}
Here $dS_{sample}/dt$ is the entropy production rate (EPR) within the self-organized dissipative structure (SODS), so here and everywhere $dS_{sample}/dt=dS_{DODS}/dt$.

The function $dS_{sample}/dt$ is a non-monotonic function of $R_{sample}$ with a maximum at the resistance matching point, $R_{sample}=R_{series}$. Based on this, we can distinguish two stages of the evolution. In the first stage $R_{sample}>R_{series}$ and in the second stage $R_{sample}<R_{series}$. The direction of the evolution is predominantly to decrease $R_{sample}$. In the first stage the power dissipated by the same increases with time (as long as $R_{sample}$ goes down). In the second stage of the evolution, when $R_{sample}<R_{series}$, the power dissipated in the sample decreases (even though the sample resistance is still going down) and the power dissipated in the resistor increases, so the total dissipated power (and the global EPR) still increases and approaches the calculated $\max(dS_{global}/dt)$ defined in Eq.\ref{eq:GlobalEPRMax}. 

Note again that, to simplify the discussion, we assume the sample's temperature equals the room temperature. This is justified because the currents flowing through the sample are usually very small, of the order of $\mu A$, which corresponds to a low level of heating in general.

Figure \ref{fig:twochannels} shows the circuit used to study the competition effect. It involves two separate samples connected in parallel. As before, there is a limiting resistor, $R_{series}$, in series with the voltage source. The resistor models the effect of scarcity or limited supply of resources available for the self-organized dissipative structures (SODS). In principle, each sample can develop a SODS, but typically, as will be discussed below, only one sample can do so.

The two samples were always prepared using the same procedure (as discussed above) and nominally identical Ag-particle concentration. The Ag flakes were placed in isopropanol and were shaken until the suspensions appeared visibly homogeneous before the measurements. No clustering was observed. The electrodes had nominally the same geometry and separation in the two samples. Thus, the two samples were designed to have statistically equivalent macroscopic initial conditions. However, the instantaneous microscopic distribution of Ag flakes cannot be expected to be identical in two suspensions. The exact positions and orientations of the flakes constitute uncontrolled microscopic fluctuations which might determine which sample eventually provides the most efficient path for the current and dissipation/entropy production. 

One of the electrodes of each sample is linked to the series resistor, while the other electrode is connected to ground through a much smaller (10k$\Omega$) resistor. This much smaller resistor is used to monitor the current for each sample. Using a high-precision, high-speed Picoscope 3403D voltmeter, the voltages on these 10k$\Omega$ resistors ($V_a$ and $V_b$) are measured using the picoscope and recorded as functions of time, using LabView. The current in the top sample (sample a) is calculated as $V_a(t)/10k\Omega$, and the current in the bottom sample (sample b) is $V_b(t)/10k\Omega$. 

By connecting our samples in parallel, we are able to quantitatively study the impact of the competition between the two self-organized dissipative structures, which potentially can develop in each sample. The samples are set to compete for the limited resources (the voltage and the electrical current). The competition occurs since as one sample self-assembles and increases its electrical conductance, the voltage on both samples drops, simply because they are connected in parallel and therefore they have the same voltage applied to them. In such a scenario, the second sample, which is a little slower and less efficient in the process of the entropy production or heat dissipation, might not be able to self-assemble at all, just because the voltage happens to become too low for self-assembly to occur, if the first sample develops a low resistance compared to $R_{series}$.

\begin{figure}[t]
    \centering
    \includegraphics[width=1\linewidth]{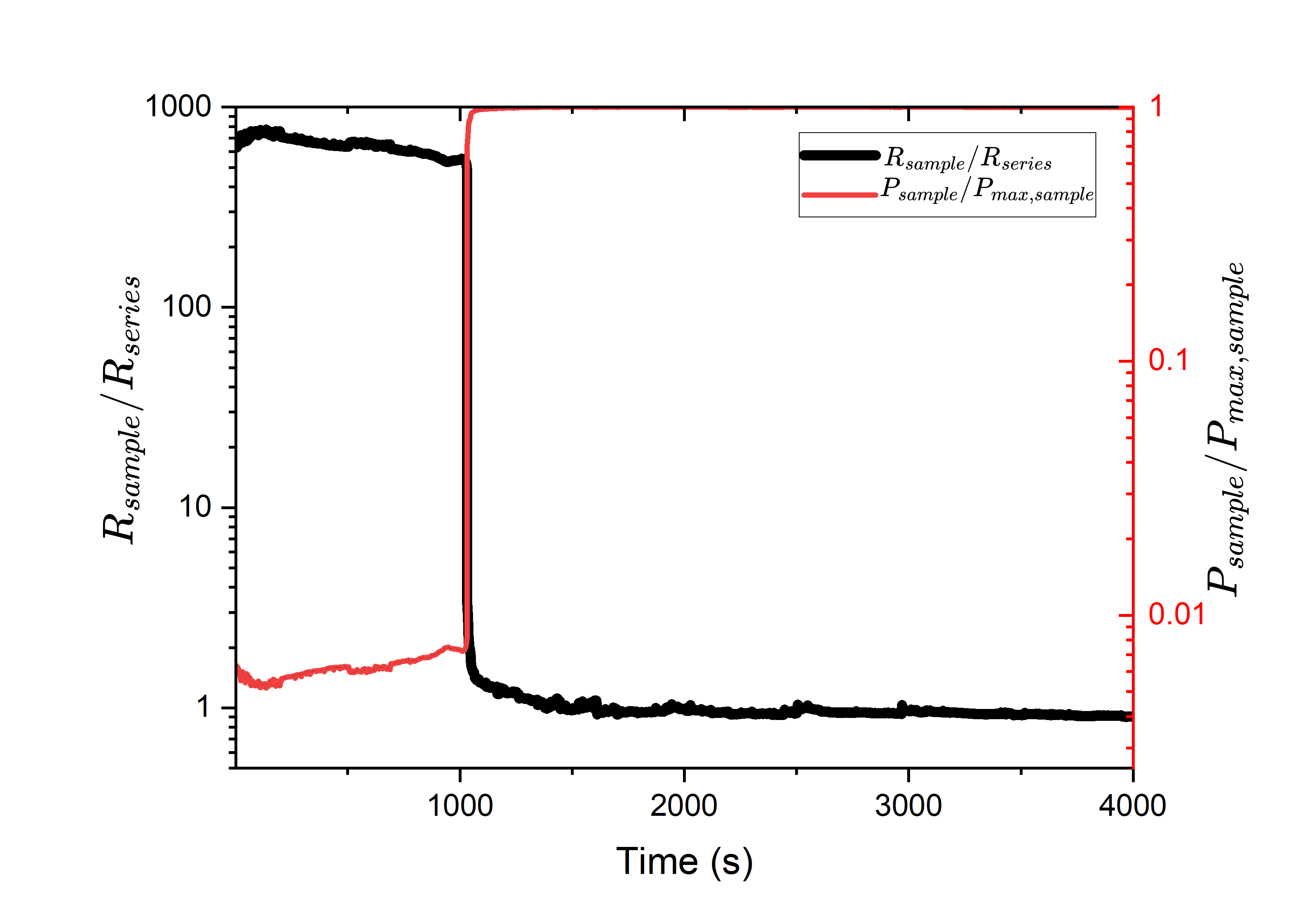}
    \caption{Measurements of the dissipative structure evolution (sample S1). The corresponding image is shown in the Fig.3.  The black curve shows the sample resistance, normalized by the value of the series resistor, plotted as a function of time. Initially this ratio is of the order of 1000, but it drops to the level of 1 in the course of self-organization of the Ag nanowires into an electrically conducting bundle, shown in Fig.3. The red curve represents the power dissipated by the sample normalized by the corresponding maximum power, $P_{max,sample}=V_{source}^2/4R_{series}$.}
    \label{fig:Agnanowirechain18Vmeasure}
\end{figure}

\begin{figure}
    \centering
    \subfigure[]{
    \includegraphics[width=0.65\linewidth]{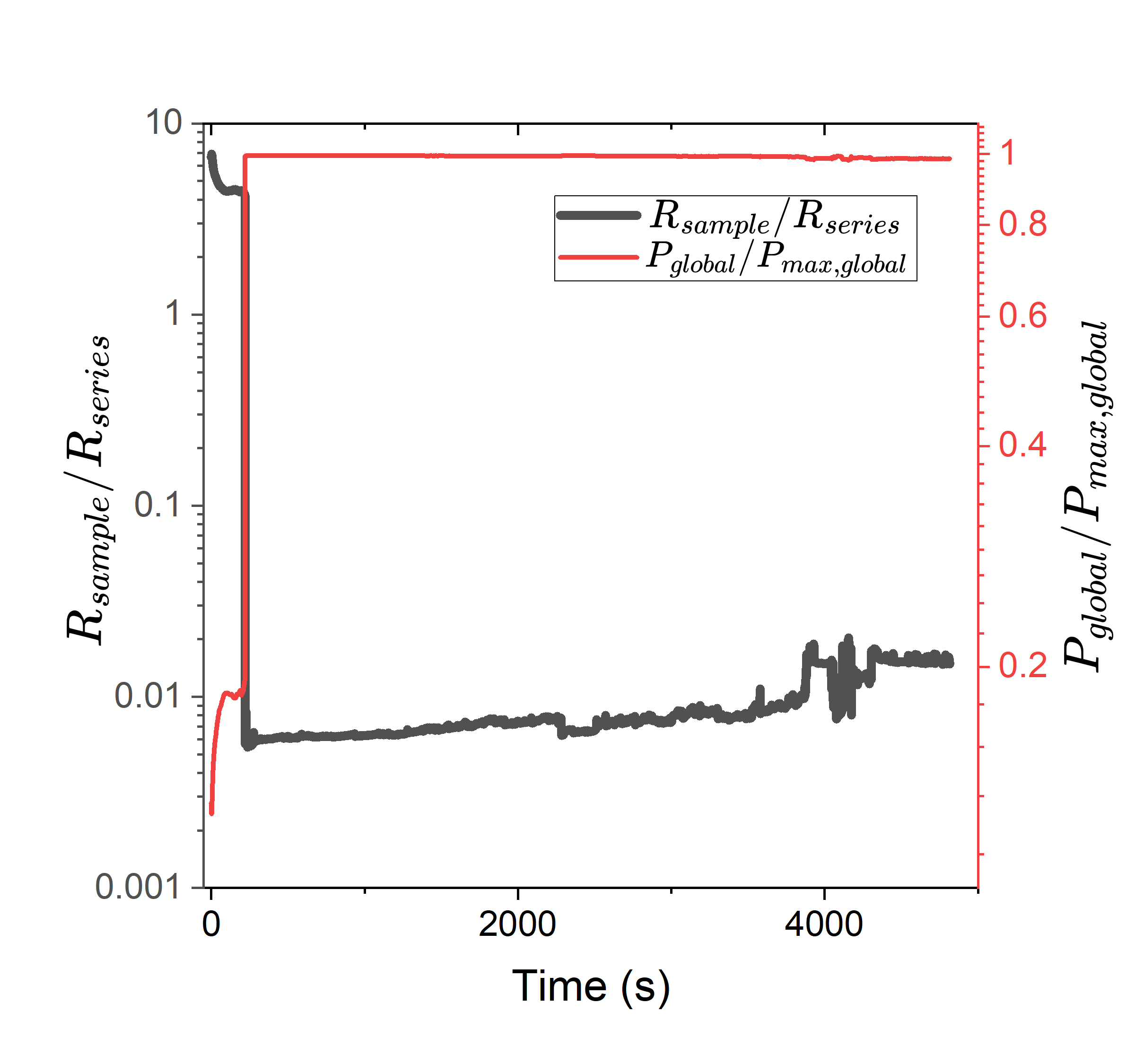}
    }
    \hfill
    \subfigure[]{
    \includegraphics[width=0.3\linewidth]{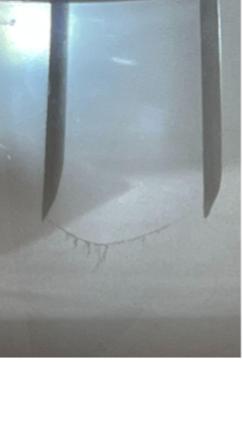}
    }
    \caption{Single-sample experiment: Evolution of Ag flakes. An Ag flake sample with a concentration of $2.87 \frac{mg}{ml}$ was run in series with a 500k$\Omega$ resistor, while the voltage bias was 70 V. (a) Sample resistance (black curve) is plotted versus time. Normalized power dissipated globally (in the resistor and in the sample) is plotted versus time (red curve). (b) An image is shown of the sample at $t =$ 4750s. A small, arched dissipative structure has formed, which connects the two electrodes. }
    \label{fig:IsopropanolFlakeEvolution}
\end{figure}

\section{Results}
 
\subsection{Single-Sample Experiments}
Single-sample measurements on Ag nanowires and Ag flakes suspensions were performed to verify their self-assembly abilities and to serve as a baseline for the following two-sample measurements. An example image of self-assembled dissipative structure (composed of Ag nanowires), namely a bundle of nanowires, is shown in Fig.\ref{fig:Agnanowirechain18V}. The source voltage was $V_{source}=18$V and the series resistor was $R_{series}=15.9$k$\Omega$. The corresponding electrical measurements are shown in Fig.\ref{fig:Agnanowirechain18Vmeasure}.  The black curve represents the normalized sample resistance normalized by the series resistor. The red curve is the power dissipated within the sample normalized by the maximum power $P_{sample,max}=V_{source}^2/4R_{series}$. In this example, the evolution progressed to the point when the resistance matching condition is satisfied and the sample EPR is very close to the maximum value. Indeed, at the beginning of the evolution the normalized power is of the order of 0.01. After approximately 200 seconds the normalized power reached the level $P_{sample}/P_{sample,max}=1$ and remains at that level up to about 5000 seconds, at which point the measurement was stopped.

We have performed many such tests on Ag nanowire suspensions in isopropanol. The tested concentrations were in the range between $0.025$ and $1.5 \frac{mg}{ml}$. With such concentration the initial resistance of the suspension is much higher than the series resistor. If a sufficient voltage is applied, $V_{source} > 9V$, the suspension exhibits self-assembly and the resistance drops by a few orders of magnitude, this drastically increasing the entropy production rate (EPR). Various $V_{source}$ voltages have been tested, in the range between $9V$ and $30V$. The series resistors were in the range between $10k\Omega $ and $500k\Omega$. One significant result is that not every sample tested in these ranges was able to form SODS, even under nominally the same conditions. The limiting process is the precipitation of the suspended nanowires on the bottom of the container. It was observed that higher voltages enable a faster evolution time and thus a higher probability of the SODS formation, which has to happen before most of the Ag nanowires precipitated. One obvious conclusion from these experiments is that the EPR maximization rule has a probabilistic character.

The calculated maximum EPR cannot be achieved instantaneously. It takes some time for the SODS to develop and to become an efficient entropy producer\cite{Bezryadin-2016}. We will call this time the evolution time, $t_e$. We define the "evolution time", $t_e$, as the time interval measured from the moment when the voltage is applied to the moment when $dS_{sample}/dt=(dS_{sample}/dt)_{max}$ for the first time. This same condition is equivalent to the resistance matching condition $R_{sample}=R_{series}$. Equivalently the same condition can be formulated as $P_{sample}=P_{sample,max}$. In other words, $t_e$ is the time when the first stage of the evolution ends and the second stage begins. For example, in Fig.\ref{fig:Agnanowirechain18Vmeasure} the evolution time was about 1100 seconds. 

\begin{figure}
    \centering
    \includegraphics[width=1\linewidth]{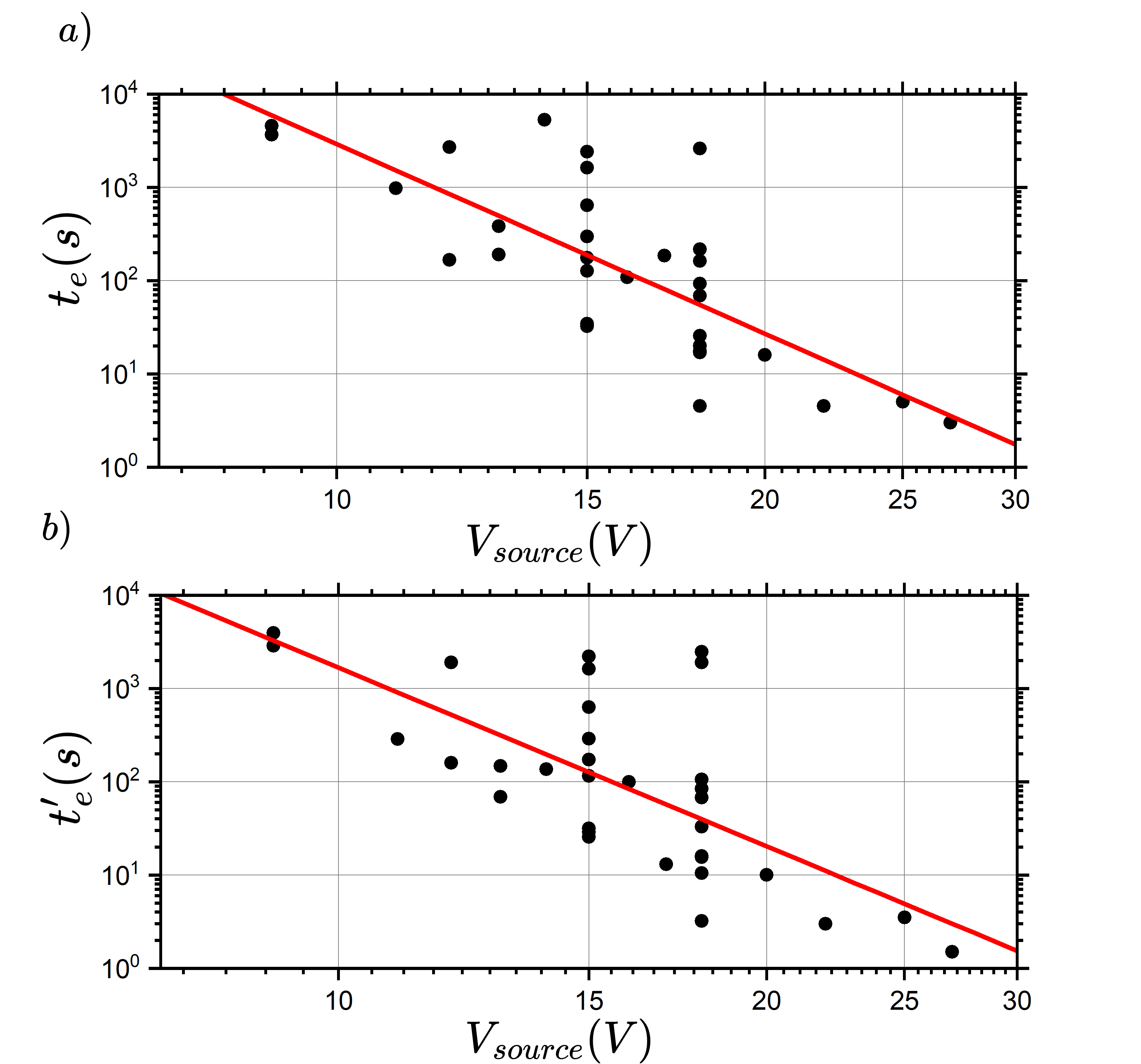}
    \caption{Two metrics of evolution time, $t_e$ and $t_e^{'}\leftrightarrow\left(\frac{dR_{series}}{dt}\right)_{min}$ are plotted versus the source voltage, $V_{source}$, for Ag nanowire suspensions in (a) and (b) respectively. All suspensions had a concentration of $0.2 \frac{mg}{ml}$ and used a series resistor with $R_{series}=$15.8k$\Omega$. The red lines show power-law fits. In (a) the best fit line shown is $t_e = 1.62\cdot 10^{10}V^{-6.75}$. In (b), the best fit line is $t_e^{'} = 3.81\cdot10^9V^{-6.36}$. Thus, the evolution time exhibits an approximate inverse power law relationship with the source voltage. }
    \label{fig:VvsEvTime}
\end{figure}

Another possible definition for the evolution time is the time at which the sample resistance exhibits a sharp decrease, or the time where $\frac{dR_{sample}}{dt}$ is at its minimum, while its absolute value is at its maximum. We label this time $t_e^{'}$. As seen in Fig.\ref{fig:VvsEvTime}, this metric behaves very similarly to $t_e$. This metric was used with samples which were not able to achieve the matching condition, but still exhibited self assembly.

\begin{figure}
    \includegraphics[width=1\linewidth]{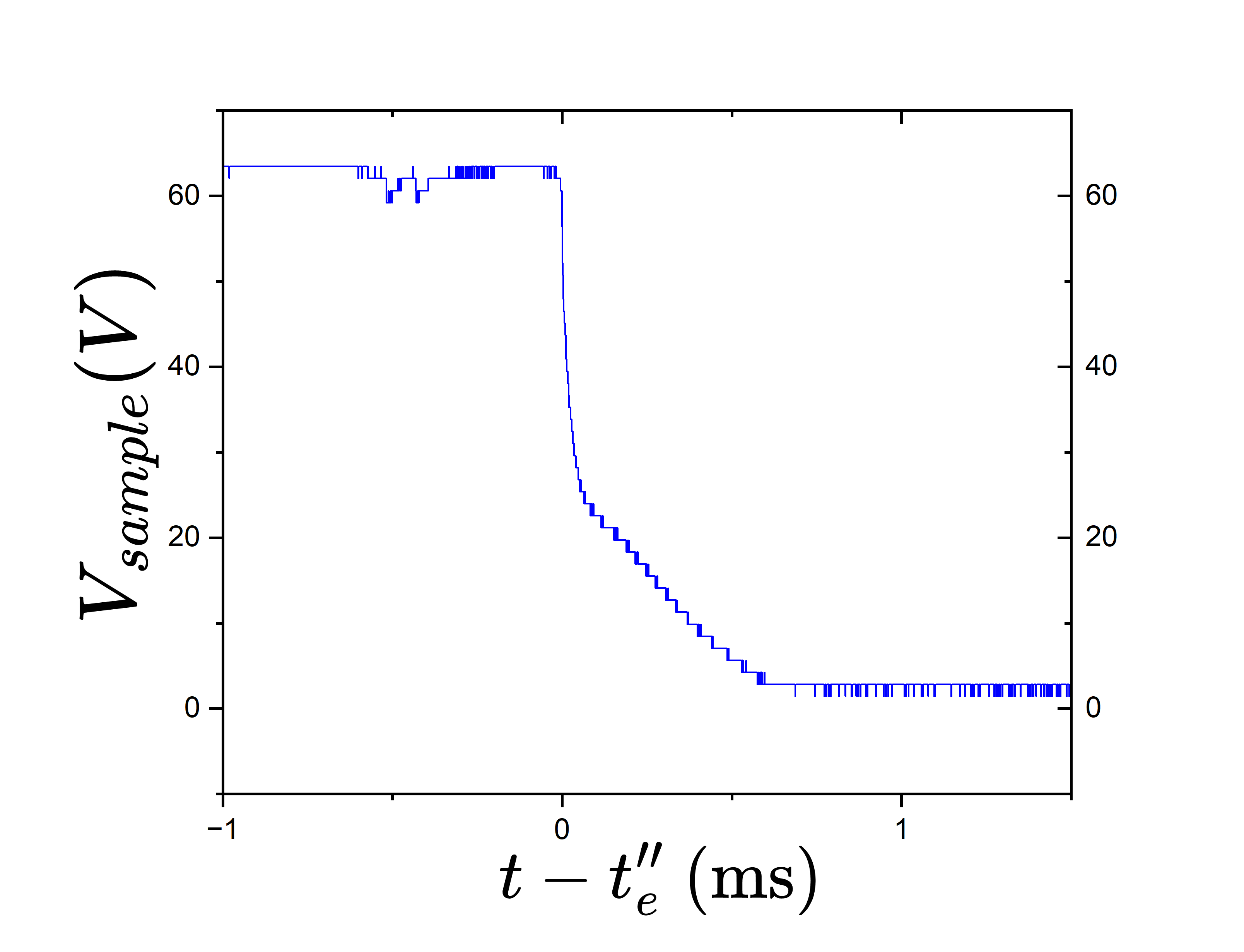}
    \caption{Single-sample experiment: Dissipative dynamic phase transition during the Ag flake self-assembly. The sample contains isopropanol with suspended silver flakes at a concentration of 0.5 $\frac{\text{mg}}{\text{ml}}$. The series resistor was 500k$\Omega$ and the source voltage was $80$V. The voltage on the sample, which is the Ag flake suspension, was measured with a high rate digitizer, Picoscope. The time axis shows the relative time synchronized to the first moment a sharp drop in  resistance is detected, $t_e''$. The self-assembly dissipative transition happens within $\Delta t \approx 0.5$ms.}
    \label{fig:Flakepicoscope}
\end{figure}

\begin{figure}
    \centering
    \includegraphics[width=1\linewidth]{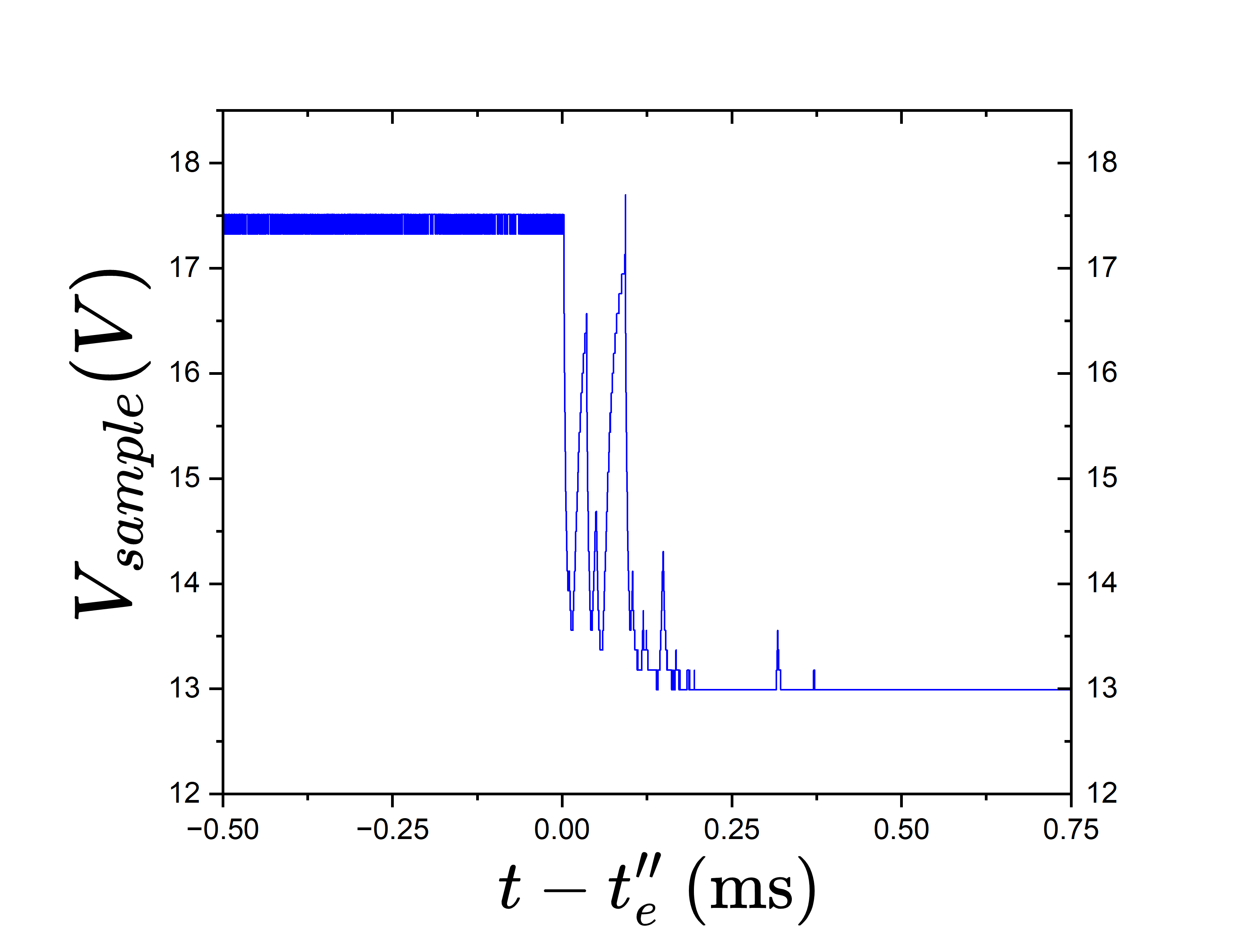}
    \caption{Single-sample experiment: Dissipative dynamic phase transition during the Ag nanowire self-assembly. A 100 nm-diameter Ag nanowire suspension, having a concentration of $0.2 \frac{mg}{ml}$ was measured in series with a 52kOhm resistor. The voltage drop across the Ag Nanowire was measured using a Picoscope digitizer, to achieve a high time resolution. A large voltage drop happens when the nanowires are able to form a continuous bundle, at $t=t_e$. The self-assembly transition occurs within $\Delta t\approx 0.25$ms The time is expressed in $t-t_e''$, where the zero is the first time ($t_e''$) a sharp drop in  resistance is detected.} 
    \label{fig:PicoScope graph 18V}
\end{figure}

Previously it was found that the evolution time has, approximately, a power-law dependence on the the voltage applied ($V_{source}$). The results have been reported in the experiments on carbon nanotubes\cite{Bezryadin-2016}. Here we confirm this phenomenon, as seen in Fig. \ref{fig:VvsEvTime}. The result is that $t_e\sim  V^{\alpha}$, where $\alpha\approx -6.5$. We also observe that the evolution process is stochastic, as for the same voltage values, the parameters $t_e$ can vary up to 2 orders of magnitude. In these experiments, some of the samples were newly prepared while some were re-agitated by manual shaking of the glass container. No difference was identified between these two approaches. 

\begin{figure}
    \includegraphics[width=1\linewidth]{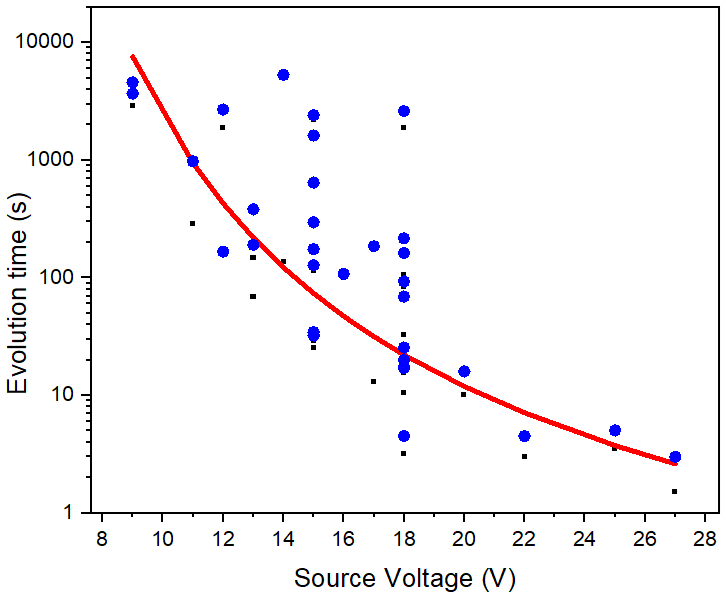}
    \caption{Single-sample experiment: Evolution time comparison to the combine activation-law and exponential-decay formula $t_1=T_{ds}\exp (V_b/V_{source})\exp(-V_{source}/V_2)$. The fitting parameters are the SODS formation time parameter $T_{ds}$=0.15s, the evolution barrier magnitude $V_b=100$V, and the evolution-time decay scale $V_d=32$V. The experimental evolution time $t_e$ is shown by blue circles and $t_e^{'}$ by black squares. The data are the same as in Fig.\ref{fig:VvsEvTime}}
    \label{fig:ExpActiv}
\end{figure}

It turns out that combined  exponential and activation functions can also be used to approximate the evolution time. First, we observe that a simple exponential dependence of the form $t_e\sim exp(-\alpha V_{source})$ does not make sense since at zero voltage it predicts a finite evolution time while one naturally expects an infinite evolution time if the driving force ($V_{source}$) is zero. Therefore we propose a function of the form $t_e\sim \exp(-\beta V_{source})\exp(\gamma/V_{source})$, which combines an exponential high rate ($\exp(\beta V_{source})$) of the nucleation of SODS at high voltages and the activation behavior at low voltages. The activation law means that the SODS nucleation rate is very low ($\exp(-\gamma/V_{sourc})$) at low driving force. Note also that the rate of nucleation of SODS is inversely proportional to the evolution time.  Qualitatively, the formula proposed above makes sense since at high driving force the evolution happens very quickly while at low driving force (source voltage) the evolution time is extremely long, which is reflected by the exponential activation factor. The activation factor suggests that to form a dissipative structure the system has to overcome a certain free energy barrier. A comparison of this combined activation law and exponential decay formula to the data is shown in Fig.\ref{fig:ExpActiv}. The fit captures the general trend, although the fluctuation are strong, indicating that SODS formation is a stochastic process which is not guaranteed to occur with certainty.

\begin{table}[htbp]
\centering
\caption{Summary of multiple tests of the evolution efficiency in single-sample experiments with Ag flake suspensions. The final current, $I_{final}$, measured after the sample stabilizes, is normalized by the theoretical maximum current $I_{max}=V_{source}/R_{series}$. The source voltage tested ranged from 45 V to 100 V. Either 100 k$\Omega$ or 500 k$\Omega$  series resistor was used. Additionally all measured samples had a concentration of $2.87\,\frac{\text{mg}}{\text{ml}}$ and all formed a SODS. The mean value for the normalized current is 0.966 and the standard deviation is 0.033.}
\label{tab:single_sample_ag_flake}
\begin{tabular}{lcc}
\toprule
$V_{source}$ & $R_{series}$ & $I_{final}/I_{max}$ \\

45V  & $500k\Omega$ & 0.854 \\
50V  & $500k\Omega$ & 0.968 \\
60V  & $500k\Omega$ & 0.945 \\
70V  & $500k\Omega$ & 0.986 \\
70V  & $500k\Omega$ & 0.986 \\
70V  & $500k\Omega$ & 0.986 \\
80V  & $500k\Omega$ & 0.994 \\
80V  & $500k\Omega$ & 0.969 \\
80V  & $100k\Omega$ & 0.958 \\
80V  & $100k\Omega$ & 0.945 \\
80V  & $100k\Omega$ & 0.957 \\
80V  & $100k\Omega$ & 0.954 \\
90V  & $500k\Omega$ & 0.989 \\
90V  & $500k\Omega$ & 0.988 \\
90V  & $500k\Omega$ & 0.972 \\
90V  & $100k\Omega$ & 0.940 \\
100V & $500k\Omega$ & 0.987 \\
100V & $500k\Omega$ & 0.980 \\
100V & $100k\Omega$ & 0.997 \\
100V & $100k\Omega$ & 0.974 \\
100V & $500k\Omega$ & 0.962 \\

\end{tabular}  
\end{table}

Experimentally, we also preformed numerous tests of SODS formation in single-sample Ag flake experiments with varying experimental parameters according to Table \ref{tab:single_sample_ag_flake}.

 In comparison to the tests on Ag nanowires, the Ag flakes have a lower resistance on average and thus they can produce structures approaching the global maximum of EPR. A self-assembly of Ag flakes into a SODS structure was observed, similar to Ag nanowires. An example of the self assembly is shown in Fig. \ref{fig:IsopropanolFlakeEvolution}. In this case also, the sample resistance remains very high for $t<t_e$, and then it sharply decrease by about $2$ orders of magnitude. The evolution time $t_e$ is about 300s in this example. At $t=t_e$ a chain of Ag flakes finishes its formation and becomes continuous. Thus the resistance drops to a level much lower than $R_{series}$. It is a typical situation for our samples with Ag flakes that $R_{sample}(t)\ll R_{series}$ for $t>t_e$. The local entropy production (i.e., within the sample itself) increases for a short time (during the transition) and then drops to a very low level because the voltage on the sample decreases to a very low level if $R_{sample}\ll R_{series}$. At the same time, the global entropy production approaches the theoretical maximum very closely, i.e., $dS_{global}(t)/dt\approx(dS_{global}/dt)_{max}$ for large times $t$. This experiment reveals a significant contrast between the entropy generation within the sample, which actually gets minimized after the evolution is complete, and the global entropy production which is maximized. This is clearly illustrated in Fig. \ref{fig:IsopropanolFlakeEvolution} where the normalized power dissipated globally (red curve) starts at a low level and approaches unity at time larger than the evolution time (which is about 300 s). The normalization factor is the maximum possible global Joule heating power $P_{max,global}=V_{source}^2/R_{series}$. It is proportional to the theoretical maximum entropy production rate as $(dS_{global}/dt)_{max}=P_{max,global}/T$.

The phase transition from almost no entropy production to a high level of entropy production happens extremely fast for both Ag Nanowire and Ag flake samples. For Ag Flakes, this dynamic phase transition has been investigated using a fast voltage digitizer (picoscope) and the results are shown in Fig.\ref{fig:Flakepicoscope}. There the voltage on the sample is plotted versus time. The $V_{sample}$ vs. $t$ curve is shown near this dissipative phase transition, i.e., for $t$ near $t_e''$. Here $t_e''$ is defined to be the first instant we observe a significant decrease in the sample resistance. A similar curve is shown for Ag nanowires in Fig. \ref{fig:PicoScope graph 18V}. The transition time, $\Delta t$, is observed to be on the order of one millisecond, which is three orders of magnitude smaller than the fastest observed evolution time, i.e., $\Delta t\ll t_e$. This difference in timescale makes $t_e' \approx t_e''$ when considering the general evolution of the SODS. An interesting observation is that the change of the voltage is non-monotonic. This shows that the entropy production rate can depend non-monotonically on time. 

Ag nanowires and Ag flakes showed similar types of behavior under the voltage bias. However, using similar concentrations and series resistors, Ag flakes required a higher voltage to self-assemble. Voltages between $40-300V$ were required to observe a self-assembly and an EPR increase, while lower voltages showed no self-assembly. Samples were measured under the voltage bias until all Ag particles precipitated, after which no self-assembly was possible. This suggests that the lack of self-assembly was not due to the time the experiment was run. A reasonable explanation for this difference is the increased size of Ag flakes. Ag flakes used had $\sim20\mu m$ diameter, while Ag nanowires used had $\sim100nm$ diameter and $\sim150\mu m$ length. A rough estimate suggests that Ag flakes would weigh about $\sim10^{-8} g$ per particle, while Ag nanowires would weigh about $\sim 10^{-11} g$ per particle. The gravity effect is the factor which competes with self-assembly and it makes it less probable for the heavier Ag flakes to form a SODS. A conclusion from this is that the EPR maximization principle might be valid only if the driving force is sufficiently strong to overcome the competing destructive effects, such as thermal fluctuations and the gravity in our model systems.

This dissipated energy is very low if $R_{sample}\gg R_{series}$. This is the case at the beginning of the first stage of the evolution when there is no SODS and all metallic particles are disorganized. Also, the sample EPR approaches zero as $R_{sample}\ll R_{series}$. This is the case at the end of the second stage of the evolution, if the evolution was successful and Ag particles were able to organize into a chain with high electrical conductivity. In the second stage of the evolution, i.e., when $R_{sample}<R_{series}$, the EPR increases globally, while the EPR generated by the SODS decreases as the structure becomes more and more organized and its resistance drops. In the second evolution stage the entropy is mostly produced by the series resistor.

The global EPR, $dS_{global}/dT=(1/T)IV_{source}$, is proportional to the current since the voltage of the voltage source is fixed.  Typically we find that throughout both evolution stages the resistance decreases with time, the current increases and therefore $dS_{global}/dt$ increases. If the evolution is fulfilled completely then at the end of the second evolution stage the sample resistance is negligible compared to the series resistance. Thus the global entropy production approaches its calculated absolute maximum, which is give in Eq.\ref{eq:GlobalEPRMax}.
This absolute maximum is never achieved exactly since its realization requires zero resistance of the SODS. (Zero resistance of the dissipative structure might be possible if the particles are superconducting but this is obviously not the case in the presented experiments.)

\begin{figure}

    \centering

    \subfigure[]{
    \includegraphics[width=0.45\linewidth]{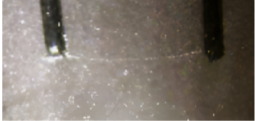}
    }
    \hfill
    \subfigure[]{
    \includegraphics[width=0.45\linewidth]{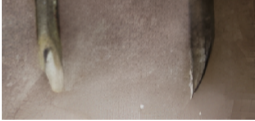}
    }

    \caption{Experiment 1. Images of the two samples obtained at the final stage of the evolution, i.e., when the global entropy production rate was near its absolute maximum. Two-parallel-samples measurements of the currents versus time. $2.87 \frac{mg}{ml}$ Ag flakes were run with a $500k\Omega$ series resistor and an applied voltage of $80V$. Image a) shows the red sample, and image b) shows the blue sample. A visible SODS is observed to form in sample a), which is associated with a high entropy production, while sample b) has a low entropy production, and there is no visible dissipative structure. The difference between samples a) and b) shows the effect of competition of the MEP. A higher entropy production would be possible if both samples formed dissipative structure, but due to the competition between the samples, only one sample is able to form a SODS.}
    \label{fig:TwoChannel}
\end{figure}

\begin{figure}
    \centering
    \includegraphics[width=1\linewidth]{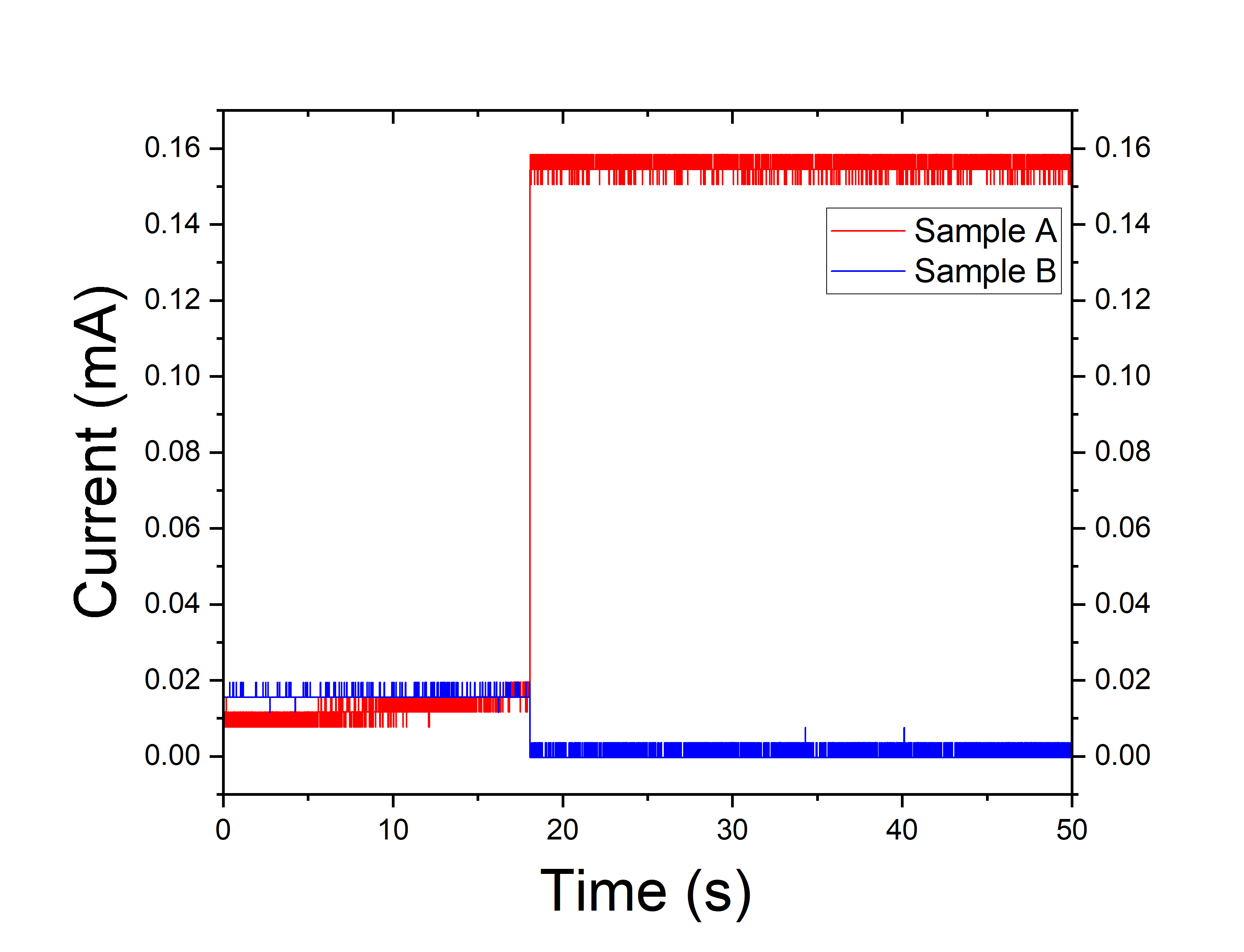}
    \caption{Experiment 1: Competitive self-assembly in Ag flakes. Two suspensions of Ag flakes at $2.87\frac{mg}{ml}$ were connected in parallel. Also, a series resistor of $R_{series}=500$k$\Omega$ was included in series with the voltage source ($V_{source}=80$V), to set up a maximum total current limit (0.16mA). At $t= 20s$, sample A is observed to self assemble, corresponding to a large increase in current (red curve). However, at the moment when the current in Sample "A" is increased the current in sample "B" drops to near zero and stays near zero permanently.}
    \label{fig:twosamples}
\end{figure}

\begin{figure}[t]
    \centering
    \includegraphics[width=1.05 \linewidth]{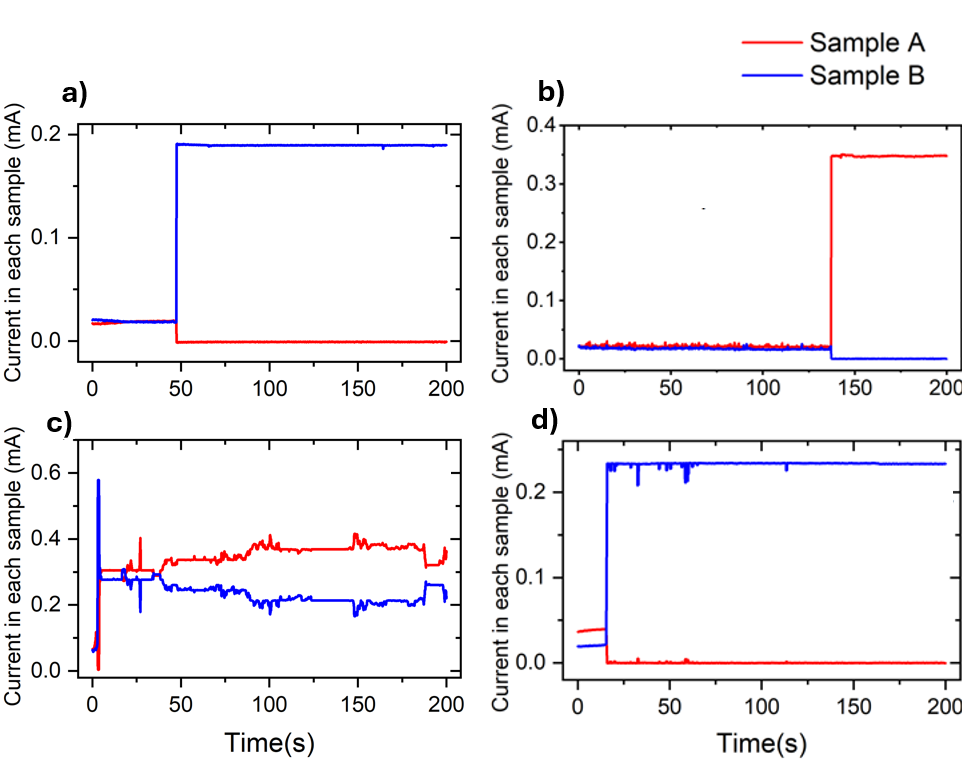}
    \caption{Examples of two-sample  Ag flake competition where the samples are denoted as sample A ("red") and B ("blue"). The voltages were 100 V in (a), 180 V in (b), 300 V in (c) and 120 V in (d). The concentrations of Ag particles in ethanol were between $1.7\frac{mg}{ml}$ and $3.7 \frac{mg}{ml}$. In all cases, the voltage source is connected in series with a resistor $R_{series} = 500 k\Omega$. A symmetric moving average over 100 individual measurements of the current is shown. Almost all samples showed self-assembly in only one sample. At high voltage and concentrations, the exclusion was not complete, as, for example, in (c) the sample "B" shows some nonzero current. Even in this case, as the current in sample "A" increases, the current in sample "B" goes down, i.e., a competition effect is still present.
    }
    \label{fig:10 Ag Flake summary}
\end{figure}

\begin{figure}
    \centering
    \includegraphics[width=01\linewidth]{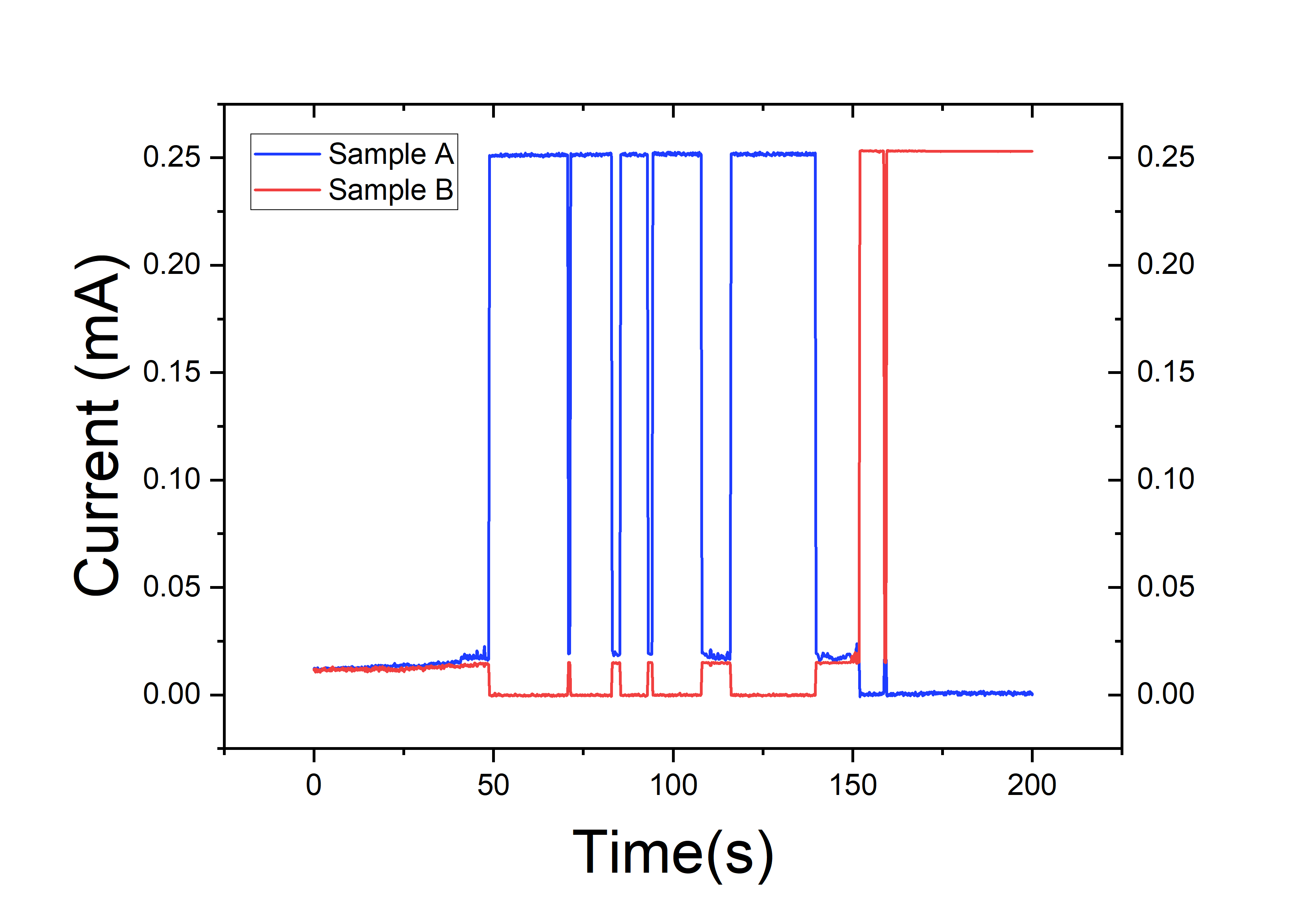}
    \caption{"Winner"-"looser" swapping in a two-sample competition experiment performed with suspensions of Ag flakes. Two Ag flake samples were connected in parallel. Both samples were prepared at a concentration of $1.72 \frac{mg}{ml}$. The voltage source ($V_{source} = 130V$) is connected in series with a resistor $R_{series}=500k\Omega$. Each curve represents a symmetric moving average over 100 individual measurements of the current in each sample. The two samples exhibited a competitive behavior. The sample A is able to form a SODS first; however, at $t \approx 140-150$, the conducting (dissipative) structure  in the sample A breaks completely. At the same time, sample B is able to form a conducting structure and generate entropy.}
    \label{fig:Ag Flake Competition}
\end{figure}

\subsection{Two-Sample Competition Experiments}

The main focus of this paper is on the competition concept. Competition in biology is understood as an interaction between organisms or species in which both require resources, the supply of which is limited. Examples of such supplies include food, water, territory, and sunlight\cite{begon1996}. According to Lang and Benbow, competition reduces the fitness of the organisms involved. This is because the functioning of one organism reduces the amount of the resource available to the other. \cite{Lang-2013}

Our experiments are designed to demonstrate and quantitatively investigate the competition effect in an artificial system (an electrical circuit) with the ability of self-organization. The competition effect is observed if two samples are connected in parallel. Each sample contains replicate identical suspensions of Ag particles, and each sample is equipped with a pair of electrodes as in Fig.\ref{fig:twochannels}. The physical implication of the parallel electrical connection is that the voltage drops on both samples are the same at all times (Fig.\ref{fig:twochannels}). Therefore, if one sample produces an SODS, its resistance drops, and the voltage on both samples decreases sharply. The corresponding circuit diagram for the two-sample experiments is shown in Fig.\ref{fig:twochannels}.
This electrical circuit also proved a model system in which resources for SODS survival are limited. Such a limit is achieved by including a resistor in series with the voltage source. Therefore, the electrical current, which is the "resource" for which the samples "compete," is limited. The largest possible current is $I_{max, total}=V_{source}/R_{series,+}$, where $V_{source}$ is the bias voltage provided by a Keithley 6517B, and $R_{series,+}=R_{series}+5k\Omega$  is the resistance after including the current limiting resistor ($R_{series}$) in series with the total resistance of two 10k$\Omega$ resistor in parallel (5k$\Omega$). 

\begin{table}[htbp]
\centering
\caption{Summary of experimental parameters and the normalized current, $I_{final}/I_{max}$ through the "winner" sample, in two-sample Ag flake experiments. The last line shows two numbers since both sample showed a significant current. In all other cases the final current through one sample (the "winner" sample) is strong, while the current through the other sample is negligibly low. The average value for the normalized current is $\langle I_{final}/I_{max}\rangle= 0.976$. The standard deviation for $I_{final}/I_{max}$ is 0.015. For $V = 300V$, both samples displayed self assembly, therefore both values of $I_{final}/I_{max}$ are listed, while for other runs only one of the samples developed SODS.}
\label{tab:two_sample_ag_flake}
\begin{tabular}{lccc}
\toprule
$V_{source}$ & $R_{series}$ & Concentration $\left( \frac{\text{mg}}{\text{ml}}\right)$ & $I_{final}/I_{max}$ \\
80V  & $500k\Omega$ & 2.87 & 0.977 \\ 
80V  & $500k\Omega$ & 2.87 & 0.980\\ 
80V  & $500k\Omega$ & 2.87 & 0.994\\
80V  & $500k\Omega$ & 2.87 & 0.986\\
80V  & $500k\Omega$ & 2.87 & 0.951\\
80V  & $500k\Omega$ & 2.87 & 0.981\\
80V  & $500k\Omega$ & 2.30 & 0.968\\
100V & $500k\Omega$ & 3.73 & 0.946\\
100V & $500k\Omega$ & 2.30 & 0.952\\
120V & $500k\Omega$ & 2.87 & 0.980\\
125V & $500k\Omega$ & 1.72 & 0.989\\
130V & $500k\Omega$ & 1.72 & 0.984\\
180V & $500k\Omega$ & 1.72 & 0.979\\
200V & $500k\Omega$ & 3.73 & 0.997\\
300V & $500k\Omega$ & 3.73 & 0.540+0.439=0.979 \\
\end{tabular}
\end{table} 

A typical example of such 2-sample experiment (referred to as  "Experiment 1") is presented in Fig. \ref{fig:twosamples}, where two samples, connected in parallel and containing identical Ag flake suspensions, were measured in series with a resistor $R_{series} = 500k\Omega$ while the source voltage was $V_{source} = 80V$. The electrical current in each sample is plotted versus time in Fig.\ref{fig:twosamples}. At time $t\approx19$s a sharp current increase is observed in sample "A" (red curve). At this instant, the current through sample "B" dropped almost to zero (blue curve) and remained at zero level permanently. 

In this example (Experiment 1; Fig.\ref{fig:twosamples}) the sample "A" (red curve) behaves as a "competitive dominant" since it is able to "out-compete" the other sample. In this model experiment the samples "compete" for the electrical current, which is limited by the series resistor. 

The photographs of the two samples corresponding to Experiment 1, discussed above, are shown in Fig. \ref{fig:TwoChannel}a and \ref{fig:TwoChannel}b respectively. There, the sample "A" is shown on the left and it clearly exhibits a chain of silver particles, which is the arch-like self-organized dissipative structure (SODS). The sample "B" is on the right; it does not show any structure. This observation confirms that among the two competing sample only one sample can "survive" (Fig. \ref{fig:TwoChannel}a. In the other sample the silver flakes eventually precipitate to the bottom of the container (not shown) and are not able to contribute to the entropy production (Fig.\ref{fig:TwoChannel}b). 

The sample which gets more electrical current generates more heat and entropy, either within itself or it causes and controls more entropy production in the series resistor. This is in agreement with the competitive exclusion principle, primarily known in ecology. It states that if two species compete for the same limiting resource, the species that uses the resource more efficiently will eventually exclude the less efficient competitor from that niche. From our experiment, it follows that "more efficient" physically means a higher global entropy production rate. In the example considered, sample "A" increases its electrical conductance faster than sample "B". As the conductance of one sample increases, the voltage (which is the thermodynamic gradient driving the self-assembly) on both samples drops. Yet, sample "A" maintains its level of self-organization, while sample "B" is not able to form a dissipative structure. This is because, as sample "A" achieves a high conductivity (at $t\approx19$s), the voltage becomes too low for the other sample (sample "B") to self-organize into a SODS. As a result, sample "A" (the red curve in Fig.\ref{fig:twosamples}) gets almost all of the available electrical current and wins the competition to generate more entropy.

Such 2-sample experiments (e.g., Experiment 1; Fig.\ref{fig:twosamples}) also illustrate the path selection principle proposed by Swenson\cite{Swenson1989,Swenson2025}:  The sample which provides the most efficient path is stabilized while the less efficient sample switches off. Our model systems allow us to elucidate the microscopic mechanism driving this path selection principle: Once SODS forms, it reduces the driving thermodynamic potential, thus making it less probable for competing SODS to occur. The more efficient the existing SODS is, the lower the thermodynamic potential and thus the lower the probability of a competing SODS occurring. Note that according to our 1-sample experiments (see the above), the probability of the SODS development decreases exponentially if the driving potential is reduced.

This dissipative structure competition phenomenon has been confirmed by performing experiments at various voltages and concentrations, as shown in Fig. \ref{fig:10 Ag Flake summary}. We observe, consistently, that if one sample forms a dissipative structure, the current in the second sample drops (almost to zero in most cases) and it loses the chance to form a dissipative structure, for the same reason, i.e., the reduction of the driving force (the sample voltage). 

The "winner" sample is not determined by a difference in the initial Ag concentration or temperature, since the samples are nominally identical. The two samples were always prepared using the same procedure (as discussed above) and nominally identical Ag-flake concentration. Rather, the competition is a highly nonlinear process, and it is initiated by a symmetry-breaking event or fluctuation\cite{Swenson2025}. The fact that, in most cases, only one sample develops a conducting chain confirms the path selection principle, which is the basis for LMEP\cite{Swenson1998,Swenson2020}. In our particular example, the current path is selected, either through one sample or through the other.

The path selection process in our system works as follows: The samples are connected in parallel, i.e., they are exposed to the same instantaneous voltage, by design. Initially, both samples are highly resistive, and they both experience the source voltage, $V_{source}$. This voltage bias creates an electric field, which polarizes the particles and therefore promotes formation of conducting chains\cite{Bezryadin-1999}. A small stochastic difference allows one sample to nucleate a conducting structure slightly earlier or to increase its conductance faster. Once this occurs, the electrical feedback inherent in the parallel circuit strongly amplifies that difference: as the resistance of the sample with a conducting structure decreases, causing the voltage across both parallel samples to decrease, thereby suppressing the ability of the second sample to form its own dissipative conducting structure. 

\textit{Thus, we propose a general mechanism for the path selection, competition, and natural selection: The more efficient dissipative structure suppresses the available gradients and thus prevents its competitors from evolving and dissipating thermodynamic potentials or gradients, which would be available otherwise.}

A more complex behavior has been observed at very high voltages, which exemplify the case of almost unlimited resources. In such cases both samples can form sustained dissipative structures , as shown in Fig. \ref{fig:10 Ag Flake summary}c. The limiting current, $I_{max}$, corresponding to Fig. \ref{fig:10 Ag Flake summary}c is 0.594 mA, which is larger than in the other examples. Therefore SODS form in both samples, but a competition effect is still present: We observe that when one structure becomes more conducting (exhibits a higher EPR) the other sample becomes less conductive (exhibits a lower EPR), and they alternate between these two regimes. Thus, \textit{at sufficiently large resource availability, the competition becomes too weak to suppress the self-organization in the less efficient sample completely. Competition is resource-dependent: when the driving resource is insufficient to sustain two dissipative structures, competition produces winner-takes-all behavior; when the driving resource (or a gradient) is sufficiently abundant, both structures can be sustained, but each time the dissipation caused by one SODS is increased, the dissipation due to the parallel SODS drops.} (Note that when both SODS are highly conducting the dissipation caused by each sample takes places in the resistor to a large extent and the dissipation caused by sample $n$ can be approximated as $P_n\approx R_{series}I_n^2$, where $I_n$ is the current in the sample $n$, and $n=1$ or $2$ in our case. Thus each sample acts as an enabler of a current but as it develops a high conductivity it shifts the dissipation to the energy supply circuit.) 

At a lower probability, we sometimes also observe that the sample with the dissipative conducting structure can swap during the self-assembly process, but only one dissipative structure was active at a time. An example of this can be seen in Fig. \ref{fig:Ag Flake Competition}, where the dissipative structure swaps a few times from sample A (blue) to sample B (red) and back. This switching behavior also suggests that the two samples are in competition, as the development of the sample A dissipative structure prevents the sample B from sustaining its dissipative structure and vice versa. In essence, the samples compete for the possibility of increasing global entropy at the fastest rate possible.

An interesting question is why the most efficient dissipative conducting structure deteriorates thus allowing its competitor in the other sample to develop. Deterioration of dissipative conducting structures have been observed and discussed previously\cite{Bezryadin-1999}. The proposed mechanism is a thermal rupture or Joule-heating-induced breakdown. Even though Ag particles are highly conductive, two particles touch through a tiny contact area. The constriction/contact resistance can therefore be significant. The local temperature increase can cause a solvent expansion which can disrupt the contact. Thus such local heating can cause significant increases of the contact resistance due to increased local thermal fluctuations. It was also observed previously that the formation an rapture of the chains causes an avalanche behavior\cite{Bezryadin-1999}. An evolution model, based on the principle that sharp increases of the local dissipation (entropy production) can destroy dissipative structures and cause avalanches has be presented in Ref.\cite{Bezryadin-1999}

We have preformed many two-parallel-sample Ag-flake measurements with varying experimental parameters. The results are listed in Table \ref{tab:two_sample_ag_flake}.  These results should be compared to Table I, where analogous results are shown for the 1-sample experiments. In both cases, when SODS is fully developed, the current tends to be very close to the maximum possible current. Thus, the global EPR (samples plus the energy supply circuit) approaches its calculated theoretical maximum. The local EPR (just within the samples) is very low, also in both types of experiments.

The experiments with the two competing samples show a lower standard deviation (0.015), compared to 1-sample experiments (0.033). This reduction might be due to statistical averaging which is more significant in 2-sample experiments. The difference of the normalized final current in 1-sample experiments and 2-sample experiments is very small, i.e., 0.966 versus 0.976 correspondingly, which is within one standard deviation. Thus the global entropy production might be slightly higher in the competition experiments but in both cases it is within one standard deviation within the calculated theoretical maximum. Interestingly, even when both samples develop SODS (see the last sample in Table II), the total normalized current is 0.979, which is almost exactly equal the average value, 0.976, obtained in the runs when only one sample formed SODS. 

One can speculate that the competition effect might prevent the global entropy production from achieving its global calculated maximum since if two dissipative structures would be active simultaneously they should be able to pass more current and generate more entropy. The reason behind such speculation is that the resistance of two parallel samples might be expected to become lower as compared to one active sample. Yet, the experiments do not show any significant difference in the final current (and the associated final EPR) in 1-sample experiments and 2-sample experiments.  Thus one concludes that the system always approaches the calculated global EPR maximum, whether a competition is explicitly imposed (2-parallel-sample case) or not imposed (1-sample cases). But in systems with two samples the total number of conducting paths which can potentially develop is larger. Such increase in the number of possibilities allows the system to approach the calculated theoretical EPR  maximum slightly closer, even though only one sample develops SODS in almost all cases. The most efficient path selection principle\cite{Swenson2010} appears to be applicable in both cases.

\section{Discussion}

\subsection{Path selection at work }
We interpret differences in the microscopic initial states as possible symmetry-breaking perturbations, while the selected sample is the one that maximizes the entropy at the fastest rate\cite{Swenson2020}. Initially, stochastic fluctuations allow conductive chains to begin forming independently in both samples. Once one sample develops a slightly more conductive pathway, its entropy production rate increases more rapidly, enabling it to dissipate the available electrical potential more efficiently and thereby deplete the thermodynamic driving force available to the competing sample. As a result, the voltage across the second sample falls below the threshold required for continued self-organization, arresting the growth of its dissipative structure. The surviving structure persists not because it was selected as a predetermined state of maximum entropy production, but because its faster local dissipation continuously reshapes the shared environment in its own favor, suppressing competing pathways. The final high-conductivity state is thus the emergent outcome of local path selection\cite{Swenson1998} acting throughout the evolution, rather than optimization toward a predefined final state.

An important part of our argument is the interpretation of fluctuations—not simply as random disturbances in an otherwise symmetrical or stable system, but as continual probes of the limits of the existing dissipative regime. Consider a system existing within an established dissipative regime. Fluctuations continually explore alternative configurations. Most fluctuations fail to establish a more effective dissipative pathway. Those fluctuations are absorbed back into the existing regime. Occasionally, a fluctuation creates a pathway with a steeper local entropy-energy gradient and therefore a higher local rate. Once such fluctuation reduces local gradients, it becomes preferentially selected. Thus, disorder becomes the exploratory mechanism through which new order becomes possible.

Thus, symmetry breaking does not require an external organizing force. Instead, it emerges from competition between pathways under the same underlying dynamical law. To construct a mathematical selection-effect model one would need to specify a relationship between the potential gradient and the local rate of dissipation. That relationship is what gives the selection effect a physical meaning. The system does not need to know what configuration is more organized. Nor does it need an external blueprint. The most dissipative pathways become preferentially established because the reduces local potentials to the lowest level possible, while such local potential are needed for alternative pathways to form. Consequently, increasing structural differentiation does not contradict the tendency toward dissipation. It is the ways that dissipation accelerates to the maximum level possible. This is the mechanism that enables LMEP to operate. Simply speaking, a local chain formation (which is an example of a symmetry breaking) can reduce local (electric) potential. Such local potential is needed for any competing chains to form. If it is reduced significantly then the existing structure (chain) becomes stable. If the local potential is to get reduced to its minimum then the local current needs to be at its maximum. And the maximum current represents the state with the maximum EPR.

\subsection{Stages of evolution in single-sample experiments}

Experiments show that as the system self-organizes it goes through two stages. Initially the metallic suspended particles (silver nanowire or micro-flakes) are disorganized and not connected to each other. Thus the conductivity of the fluid and the current passing through it is very low. Therefore, at the starting point of the evolution experiments, the entropy production rate is near zero. This is true for both $dS_{sample}/dt$ and $dS_{global}/dt$. In other words, at time zero, both EPR rates are much lower than their calculated theoretical maxima.

Next, the system undergoes an evolution in which the conducting particles orient themselves along the applied electric field and form continuous and electrically conducting chains, as seen in Fig. \ref{fig:Agnanowirechain18V}, Fig. \ref{fig:IsopropanolFlakeEvolution}b, and Fig. \ref{fig:TwoChannel}. Initially the chains are short but with time they form bundles which eventually link the electrodes and make the conductance of the sample much higher compared to the initial state. If the concentration of the metallic particles and their individual conductivities are sufficiently high, and the applied electric field is sufficiently strong to overcome destructive effects (e.g., gravity and thermal fluctuations), then the particles organize into a dissipative structure which eventually evolves to a state such that the sample EPR equals its calculated theoretical maximum.

A simple calculation shows that the maximum power dissipated in the sample occurs under the impedance-matching condition $R_{\mathrm{sample}} = R_{\mathrm{series}}.
$ At this point, the sample's entropy production rate reaches its calculated maximum, $dS_{sample}/dt=(dS_{sample}/dt)_{max}$
 as defined in Eq.~\ref{eq:SampleEPRMax}, marking the end of the first stage of the evolution. The EPR is given by $dS_{sample}/dt=IV_{sample}/T$
 with entropy generated as heat flows from the sample to the environment at temperature $T$.

There are in fact two evolution stages. In the first stage, when $R_{sample}>R_{series}$, the EPR increases both within the sample as well as globally. In the second stage, when $R_{sample}<R_{series}$, the global EPR continues to increase while the sample EPR decreases, because the voltage now drops mostly on the series resistor. As the sample achieves the maximum calculated EPR, the evolution does not stop at this point, although it might slow down since the voltage on the sample at this point is reduced by 50\%, due to the voltage drop on the resistor\cite{Bezryadin-1999}. In general, the evolution continues as the system evolves to the theoretical maximum of the global entropy production rate. The global EPR includes both the heat generated by the sample and the heat generated by the energy supply circuit. An important fact is that after the local EPR (i.e., the "sample EPR") maximum is achieved, and if the evolution continues further (i.e., the flux or the current allowed by the dissipative structure continues to increase), the global entropy production rate  continues to increase while the local EPR begins to decreases. Thus the signature of the second evolution stage is that the local EPR decreases while the global EPR continues to increase. Thus, we observe a tendency, which we suspect is quite general, that advanced dissipative structures tend to shifts the heat generation from themselves to outer elements, i.e., to  their energy supply system. In our case, at the second evolution stage, more heat is generated in the series resistor than in the self-assembled silver particle chain (see the derivation of Eq. \ref{eq:SampleEPRMax} and Eq.\ref{eq:SampleEPR}). But the self-assembled Ag particle chain acts as an enabler of such dissipation. Note that such two stages of evolution have been observed previously in Ref.\cite{bezryadin-2015} and it was also concluded that the first stage is characterized by a series of destructive events (avalanches), while the second stage was much more smooth and avalanche-free\cite{Bezryadin-1999}.

In the second stage of evolution ( $R_{sample}<R_{series}$ and so $V_{sample}<V_{series}$), the global entropy production continues to grow and approach the calculated global maximum, which is set by the energy supply circuit ($V_{source}$ and $R_{series}$) and is given in Eq.\ref{eq:GlobalEPRMax}. The global entropy production is maximized when $R_{sample}$ is exactly zero, however, this condition is never realized in practice. Therefore, as more silver particles join SODS, the systems evolves to the global EPR maximum but cannot achieve it exactly. Indeed, in the second (final) evolution stage we frequently observe $R_{sample}\ll R_{series}$ (see Fig.\ref{fig:IsopropanolFlakeEvolution}), provided that the suspension has enough silver flakes. The Joule heating power produced by the supply circuit is  $P_{outside}=R_{series}I^2$ and the heat power produced inside the dissipative structure is $P_{inside}=R_{sample}I^2$. The current, $I$, is the same in the SODS and in the resistor, and this current is enabled and controlled by SODS. All this heat goes to the environment at temperature $T\approx295K$, so the sample EPR is $S_{sample}/dt=R_{sample}I^2/T$ and the global EPR is $S_{global}/dt=R_{series}I^2/T+R_{sample}I^2/T$. At the end of the second evolution stage the sample EPR is much lower than the global EPR because the heat is mostly released in the energy supply circuit.  

The analogy between our model systems and biological organisms is noteworthy. At earlier stages of evolution, organisms primarily generate entropy within their own bodies. As evolution progresses, however, a larger fraction of entropy production shifts to external systems. Human civilization provides a clear example of this transition. With the development of technologies to harness chemical energy (e.g., fire and power plants), nuclear energy, and electrical energy (e.g., data centers), most entropy production associated with human activity now occurs outside the body, within engineered infrastructures and energy supply networks.

A similar pattern is observed in our experiments. As Ag particles evolve toward a highly conductive self organized dissipative structure (SODS), during the second stage of evolution, the entropy production shifts to the external energy supply circuit, specifically the series resistor. In contrast, at an earlier evolutionary stage the SODS dissipates energy mostly within itself.

By analogy, animals—representing an earlier stage—mainly dissipate energy through internal metabolic processes such as digestion. Humans, representing a more advanced stage, increasingly rely on external systems such as factories, power plants, and data centers, which dominate overall energy dissipation and entropy production. Consequently, the conversion of free energy (ultimately derived from the Sun) now occurs predominantly outside the human body, within industrial and technological systems. At present, the rate of entropy production in these external systems far exceeds the level of dissipation occurring within human bodies.

To characterize the evolution time scale quantitatively we introduced the "evolution time", $t_e$, which was defined as the time needed for the dissipative structure to achieve the sample EPR maximum, i.e. the time at which $R_{sample}=R_{series}$ for the first time. In other words, $t_e$ is the duration of the first evolution stage, according to the definition above. Our numerous tests showed that such evolution time decays rapidly as a function of the applied voltage, which is the evolution driving force. This dependence can be modeled either by a power law function or a newly introduced  activation-law and exponential decay formula. The results are summarized in Fig.\ref{fig:VvsEvTime} and Fig. \ref{fig:ExpActiv}.

\subsection{The Competition Effect in Two-Sample Experiments}

The main focus of this paper is on systems in which two samples are connected in parallel. In such configurations, Ag flake suspensions compete for a shared resource—the free energy supplied by the voltage/current source, which is required for the formation of dissipative structures and the associated entropy production.

Initially, when both samples exhibit high resistance, the voltage across each is approximately equal to the source voltage, $V_{\text{source}}$. However, once one sample undergoes self-assembly and forms a dissipative structure with high electrical conductance, its voltage drops sharply. Because the samples are connected in parallel, the voltage across both samples is always the same. Consequently, once an efficient SODS develops in one of the samples, the voltage across the second sample decreases as well, suppressing its ability to form a dissipative structure.

As a result, the system exhibits a symmetry-breaking effect: the formation of a dissipative structure in one sample inhibits its formation in the other, such that only one sample develops a dissipative structure under these conditions.

This is an example of a "winner takes all" scenario. This winner is the sample in which the dissipative structure evolves first. This sample takes the entire resource, which is the electric current, limited by the resistor. A strong voltage (electric field, to be more precise) is needed to form a continuous and conducting chain of Ag particles.  The voltage polarizes the particles and then they, acting as electric dipoles, tend to form chains. The current, on the other hand, allows the chain to survive. Indeed, if a break develops, the current creates a high concentration of opposite charges around the break point, which produce a force to heal the break. Once a structure forms it needs the current to sustain its functionality. If the current is limited then only one dissipative structure can persist. 

In 1-sample experiments, at the end of the second stage of the evolution, the current is very close to the calculated maximum current $V_{source}/R_{series}$. Indeed, as Table I shows, the difference between the achieved current and the maximum current is only about 3.4\% on average. Thus, the systems approaches the calculated global EPR maximum very closely. A similar conclusion applies to the 2-sample experiments with the competition effect. Indeed, from Table II one concludes that that current is less than the calculated maximum by about 2.4\%. Thus, in all cases the maximum entropy production principle appears to be obeyed quite well. If one compares the efficiency of 1-sample cases and 2-sample cases, they are similar. The 2-sample system shows a slightly higher normalized current on average. Interestingly, even when both samples form SODS (which is very unusual), the total level of dissipation is the same as in the runs in which only one sample is able to form SODS (this is typical). Thus the competition does not break the maximum entropy production rule, but it might make it statistically more accurate.

Previously, theoretical models have been proposed \cite{Bartlett-2015} in which two types of dissipative structures compete for free energy. We now report experiments in which the same type of SODS compete with each other. This competitive behavior resembles natural systems, as animals and plants must compete for free energy in terms of food and sunlight (for plants). The "winner-takes-all" behavior is the most prevalent under the experimental conditions investigated.

Our experiments also demonstrate that the "winners" and the "losers" are not always set forever. For example in Fig.\ref{fig:Ag Flake Competition} the "blue" sample develops a highly conducting dissipative structure at $t=50$, yet this structure gets destroyed at $t=150$s, when the "red" sample develops a similar (in terms of its conductance) dissipative structure. Extending the biological analogy, Fig.\ref{fig:10 Ag Flake summary}C shows that at very high voltages ($300V$), both samples are able to self-assemble. This represents the case of abundant free energy, where multiple competing SODS can co-exist. Yet, the competition effect is evident even in this case, as when one SODS develops a larger current (and dissipates more energy), the other one exhibits a reduces current and a reduced dissipation.

\subsection{Competition effect and the entropy production maximization}
 
The question of constraints in formulations of the entropy production extremum principle is somewhat difficult because of the possibility of circular reasoning. Whatever the entropy production rate is, one can always say that it is the possible maximum and that the constraints are just such that the observed EPR has to be considered as the maximum possible value. In this case, although always true, the extremum principle would have no predictive power. To avoid circular definitions, one needs an independent way to establish what is the maximum EPR (local or global), given the constraints. Thus, here we choose a model system shaped as an electric circuit. In this case one can calculate the maximum possible EPR, so it is known before the measurement begins.

A related question is whether the maximum EPR value is a constant or is it a function of time. If the EPR is a constant for a given system, then, an immediate conclusion is that EPR is not always at the maximum possible level since the EPR increases in a typical evolution self-organization process. It always takes a finite time for the dissipative structure to form. In particular, we find that the evolution time increases if the driving force (voltage) is reduced (Fig.\ref{fig:VvsEvTime}) following either a power-law or an exponential dependence. 

In formulations of proposed principles governing entropy production in non-equilibrium self-organizing systems, one can distinguish two approaches: MEP as a state-selection principle\cite{kleidon_2010} and LMEP as a path-selection principle\cite{Swenson1989}. The MEP selects the steady-state configuration that maximizes the rate of the entropy produced. There are many variations for 
MEP  \cite{Martyushev2021-Review, MARTYUSHEV201417, Bradford-2013, Dyke-2010-Theory}. Our results suggest that in most cases there is a final state steady state in which the global EPR is near its calculated maximum, while the local EPR (inside SODS) is near its minimum (i.e, near zero). It should be noted that this behavior should be interpreted as a statistical trend. 

The path-selection LMEP instead describes evolution proceeding in the direction of steepest ascent of entropy production. The term was coined by Swenson\cite{Swenson1989} as a proposed fourth law of thermodynamics. LMEP selects the path which maximizes the entropy produced at every instant. The key idea of LMEP is the selection of the most efficient path for the entropy production. Our competition experiment demonstrates the path selection in action, since the path for the electrical current which produces the entropy at the highest rate is selected. 

Another interesting question is the effect of "winner-takes-all" competition on the entropy production. Assuming that the conductivity of each Ag chain has an upper limit, $G_{1, max}$, one concludes that if the competition effect were absent and two chains were formed, then the total conductivity would be twice as large, i.e. $2G_{1,max}$. Under MEP, state-selection predicts this configuration, where the total current would be higher and more entropy would be produced per unit time. In such a simplified view, the competition effect prevents the system from reaching the global entropy production rate because the current is maximized in the winning sample, but it remains near zero in the sample that does not win the competition.  Interestingly, the current in a system with two competing samples is found to be slightly larger on average when compared to the experiments with one sample. Therefore one have to cnclude that the presence of the competation can make the winning sample somewhat more efficient, i.e., its $G_{winner,max}$ is higher than in 1-sample experiments ($G_{1,max}$), i.e., $G_{winner,max}>G_{1,max}$.
   
In the two-sample system, the competition effect arises naturally from path selection, whereas the state-selection framework seems to require an additional constraint to be consistent with the competition effect. Note that the state selection approach would require that each sample generates as mush entropy per unit time as possible, thus prohibiting the destruction of the less efficient SODS, while in fact such destruction does happen.

Based our results, we elucidate the dynamical physical mechanism enabling physical selection: A SODS or an autocatakinetic state gains a form of physical "fitness" when it establishes a pathway that channels available energy flows more efficiently than other possible paths. In systems with competing SODS, stochastic fluctuations in the initial conditions create symmetry-breaking events\cite{Swenson2025}, causing one sample to exhibit a high EPR compared to the other. As the leading sample's conductance increases, it draws a larger share of the available free energy supply.  As it develops and approaches the maximum EPR, it modifies the nonequilibrium environment by consuming available free energy and eliminating the gradients that could support alternative pathways. According to our experiments, such elimination happens because the energy production shifts from the samples to the energy supply circuit. Thus, at the level of the self-organized dissipative structures, the available supply of the free energy (voltage) becomes very low and alternative SODS cannot develop.  This creates a selection mechanism: SODS with the largest EPR persist, while competing structures are prohibited because the thermodynamic resources necessary for their emergence and persistence are suppressed by the highest EPR SODS. This selection does happens only if SODS is sufficiently efficient to shift the entropy production to the energy supply circuit. This principle should apply to other SODS, like biological organisms, for example. This describes the evolution rule: "greedy", instant-by-instant path selection produces total exclusion, rather than partial sharing, because the leading path's growth erases the resource gradient the trailing path depends on.

In the general sense the competition effect is present even within one sample, but it is not simple to quantify. Indeed, different areas of the same sample compete for the same resources (silver particles, electric field, electric current) which are needed to develop and maintain the dissipative structure. Yet, at the end of the evolution process, only a small volume of the sample is actually occupied by a functional dissipative structure, while other regions of the same sample (the regions outside of the silver particle chain) remain dissipation-free. Since the efficiency (quantified as $I/I_{max}$) of the SODS is a nonlinear function of the number of particles involved in the SODS, a concentration of available particles in one SODS is necessary to achieve the highest efficiency possible. Also, the SODS stability is ensured by the current in it, so, again, the SODS improves its stability if more particles participate in the current-carrying channel.

\subsection{ Global Implications: Kardashev scale and the Law of Maximum Entropy Production}
The Kardashev scale, $K$, classifies civilizations by their ability to harness energy at increasing global scales: planetary (Type I), stellar (Type II), and galactic (Type III)\cite{Kardeshev-1964}. If this framework is applied to human civilization, the Type I scale will be achieved when we unitize the amount of energy arriving from the Sun to Earth, the Type II civilization will harnesses the full energy output of the Sun (e.g., via a Dyson sphere or an analogous systems), and the Type III civilization would exploit the energy on the scale of their entire galaxy (Milky Way in our case). The human civilization is often estimated to be below Type I, and more specifically $K\approx$0.7, as we do not yet utilize the full energy budget of Earth, although our energy consumption has historically grown rapidly, roughly exponentially. 

From a thermodynamic perspective, biological life can be viewed as a self-organized dissipative structure (DODS), driven primarily by solar radiation and all the free energy sources it produces\cite{Lineweaver-2025}. The LMEP suggests that such non-equilibrium systems tend to evolve toward paths that increase their entropy production and approach the maximum, defined by constraints. Yet, according to the Kardashev scale, these constraints are not fixed in time, as life forms can, in principle, spread to other planets or even other star systems. This perspective may imply a tendency for an ever increasing energy utilization.

When interpreted alongside the Kardashev scale, the maximum entropy production rate principle implies that the progression of technological civilizations may reflect an increasing capacity to dissipate energy gradients (free energy sources), particularly through the transformation of concentrated energy sources such as stellar radiation into low temperature thermal radiation with high entropy. In such view, the constraints are not rigid since the spacial dimensions and the scope of the dissipative structure can increase, in principle, indefinitely. Energy consumption on the Kardashev scale is not just “progress”; it is the macroscopic enabler of higher entropy-production rates.
Intelligent civilizations, together with their technospheres, represent the most sophisticated dissipative structures yet known. Their emergence and progression up the Kardashev scale can be understood as a consequence of LMEP: at each stage, it favors pathways capable of dissipating available free-energy gradients at the fastest possible rate. Ascending Kardashev scale is therefore not optional or cultural—it is the thermodynamic trajectory for any long-lived technological biosphere. 

We propose that the maximum EPR principle is the driving variational principle that forces Kardashev ascent and might even predict specific observable consequences. Our technological civilization, if it survives its initial “entropy bottleneck” ($K\approx$ 0.7–0.8, the current human phase of rapid but possibly unsustainable dissipation), will inevitably reach full Type I ($K=1$) status within 200–500 years and Type II status ($K=2$) within 3,000–10,000 years after that.

\begin{figure}
    \centering
    \includegraphics[width=01\linewidth]{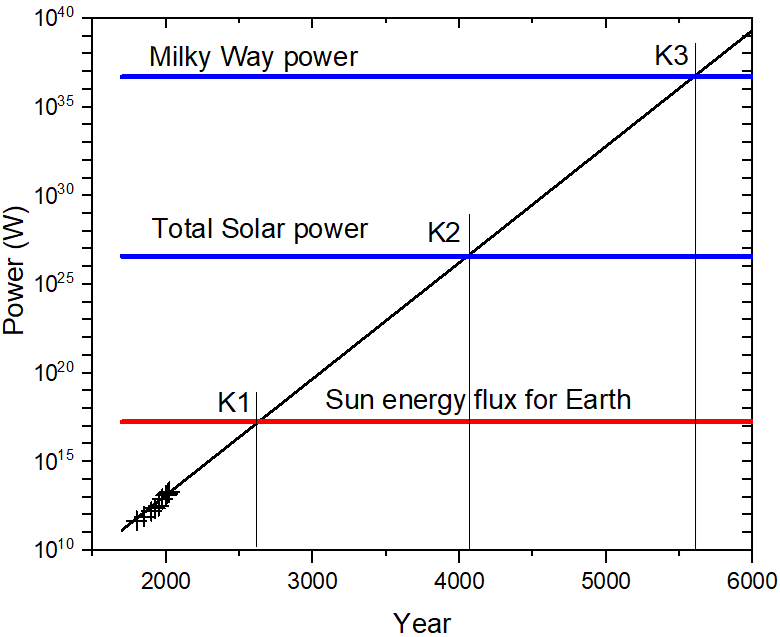}
    \caption{Maximization of the entropy production rate and the Kardashev scale. The stage-I civilization utilizes the energy available from the Sun at the Earth surface (red line). The stage-II civilization uses all free energy given out by the Sun (bottom blue line), with the help of a Dyson sphere or an analogous construct. The stage-III civilization uses all energy generated by the galaxy (top blue line). The black "+" signs show historical data of the energy consumption (per second) by the human civilization\cite{owid-energy-production-consumption}. The tilted black line is the exponential fit. Based on this graph, the corresponding critical years are 2600, 4100, 5600. Note that K1, K2, ans K3 mark the three Kardahsev levels of the power consumption.}
    \label{KS2}
\end{figure}

A projection of the energy consumption is illustrated in Fig.\ref{KS2}.  There, the black crosses show approximate power consumption by the human civilization over about two hundred years. It growth approximately exponentially and the fit (tilted black line) is $P=\text{exp}(0.015t)$, where $t$ is the time measured in years and $P$ is the power (in Watts) consumed by the human civilization. The extrapolated exponential energy consumption (tilted black line) crosses the planetary energy influx (red) line in the year $\approx$2600 and the total solar energy consumption in the year $\approx$4100. Total galactic energy consumption would happen in the year $\approx$5600, according to this simple exponential-growth model. Of course, the actual evolution is probabilistic, and these estimates might represent probable but not definite scenarios. Additionally, this estimate does not take into account relativistic effects of course, but it is solely based on the extrapolation of the current exponential growth of the energy consumption.

This timeline is consistent with LMEP: each successive self-organized state unlocks higher $dS/dt$ rates, creating a positive feedback loop and providing resources for the extensions to new free energy sources, until stellar-scale dissipation is achieved. We speculate that beyond Type II, global LMEP maximization should favor ultra-efficient, reversible-computation networks. The entropy production per bit would be as low as ln(2)$k_BT$, or possibly lower if adiabatic computation is realized, essentially approaching zero (here $k_B$ is the Boltzmann constant). Note that the heat-generating data centers of today produce much more entropy than ln(2)$k_BT$ per bit operation. Reversible computation minimizes entropy production in the computational subsystem, but LMEP cares about the global rate across the entire open system and its environment. Thus, such reversible computation allows enormous information processing (which is essentially used to discover and extend the paths for the entropy production) with minimal unnecessary waste, while the system as a whole continues to increase EPR by expanding utilization of whatever gradients (free energy) remain exploitable. Such development would “thermodynamically camouflage” itself by producing minimal detectable waste heat while still operating at the maximum possible $dS/dt$. 

Another interesting observation is that if Dyson sphere (or an analogue) is ever constructed, it would represent a shift of the entropy production from Earth closer to the Sun. Such development would reduce the flux of energy on the Earth surface. Then, evolution or even survival of other life forms on Earth would be much less probable. This is a manifestation of the competition effect also, as more efficient SODS tend to reduce available thermodynamic potentials. This is analogous to Benard convection cells reducing the temperature difference driving them or our silver particle chain reducing the voltage on it and any other samples connected in parallel with it.

The EPR growth is ultimately constrained by planetary heat dissipation. Energy consumed on Earth eventually degrades into heat that must be radiated into space. The rate at which Earth radiates thermal energy is governed approximately by the Stefan–Boltzmann law. If anthropogenic heat production were to become sufficiently large relative to Earth’s current radiative balance, it would raise global temperatures until a new radiative equilibrium was established, potentially imposing severe limits on habitability. To circumvent terrestrial thermal constraints, humanity could move increasingly energy-intensive processes, such as AI data centers, into space. Space-based radiators could then dissipate waste heat directly into the cold of space. Such efforts are already being pursued even at the time of this writing. For example, SpaceX has proposed launching up to one million satellites as orbital data centers. These would capture solar energy in space (constant sunlight, no atmosphere/weather) to power AI/computing. Musk has called this "a first step towards becoming a Kardashev II-level civilization—one that can harness the Sun’s full power."\textcolor{red}{\cite{cnn2026musk}} If Kardashev levels are taken into consideration, then it becomes clear that the main principle of the evolution of the dissipative structure is not only to maximize EPR but also to find new source of free energy so the dissipative structure expands and continues to grow indefinitely.

\section{\label{sec:conclusion}Conclusion}

We study entropy production in a nonlinear, non-equilibrium system of Ag particles capable of self-organization into dissipative structures. The energy supply circuit limits the maximum dissipation. Our systems enables a self-organized dissipative structure, and it is unique because the maximum possible entropy production rate can be easily calculated before the experiment begins.

We distinguish the entropy production within the sample (dissipative structure), $dS_{sample}/dt=(1/T)I^2R_{sample}$, and the global entropy production, $dS_{global}/dt=(1/T)IV_{source}$, which includes the sample and the energy supply circuit. The $dS_{sample}/dt$ typically can achieve its maximum value if the applied voltage is sufficiently high, although the time required for this first stage of the evolution to be completed increases rapidly if the applied voltage is reduced, roughly following an Arrhenius trend. 

In the second evolution stage (defined as after the $dS_{sample}/dt$ maximum is achieved), the conductance of the dissipative structure continues to increase, and the current approaches its calculated maximum, $I\rightarrow I_{max}=V_{source}/R_{series}$. So, in the final evolution stage, the global entropy production approaches its maximum given the constraints, yet the power dissipated within the sample approaches zero. In other words, as $t\rightarrow \infty$, we observe $dS_{global}/dt\rightarrow (dS_{global}/dt)_{max}$ while $dS_{sample}/dt\rightarrow 0$.

We also observe and quantify the effect of competition between self organized dissipative structures (SODS) in a setting where two samples are connected in parallel and thus always exposed to the same voltage, by design. In this case the two SODS compete for the electrical current, which sustains them and allows them to produce entropy. The competition is manifest in that the self assembly of a dissipative structure  in one sample usually prohibits formation of SODS in the other sample. These results mimic biological systems, where dissipative structures (organisms) compete with each other for free energy and resources, and it can be summarized as "winner-takes-all" effect. In biology and evolution, the "winner-takes-all" effect refers to situations where a small initial advantage allows one individual, genotype, species, or trait to become overwhelmingly dominant, often reducing diversity. 

 The "winner-takes-all" phenomenon is consistent with the path-selection principle 
 (LMEP). We propose a dynamical explanation for the selection process within the autocatakinetics framework. The proposed mechanism is that as a strong SODS develops, it reduces the driving thermodynamic potential gradients and thus inhibits any competitors. This reduction of the local gradients happens because the most efficient SODS shift the dissipation from themselves to the power supply circuit. The principle is probabilistic in nature.

Expanding on this biological analogy, we examine the potential implications of the law of maximum entropy production principle for human energy consumption. We suggest that human technological development and humanity’s progression along Kardashev scale is a thermodynamic trend. It is guided by the entropy production rate maximization principle.

Our experiments show that as the dissipative systems become mature they tend to shift the entropy production away from themselves. Indeed, in the first evolution stage the chains of silver particles generate heat inside themselves. Yet, in the second evolution stage the heat generation is shifted closer to the energy source, namely to the resistor in our model experiment. Similarly, in Kardashev type of evolution, the main location of the entropy production shifts from being inside biological organisms to the planetary artificial structures and then into outer space, closer to their star.

We propose the following selection mechanism for the maximization of the entropy production rate or LMEP. Given a set of free-energy sources, a nonequilibrium system capable of self-organization tends to develop competing, localized dissipative structures (SODS). Depending on their ability to shift entropy production closer to the energy sources, these structures modify the thermodynamic landscape by reducing local thermodynamic potential gradients. Consequently, more efficient SODS tend to become stable, as the local energy supplies become insufficient to sustain the development of competing SODS or alternative dissipative pathways. This constitutes the proposed selection mechanism.

\section*{References}

\vspace{-10pt}

\bibliography{references}

\end{document}